\documentclass[12pt, draftclsnofoot, onecolumn]{IEEEtran}

\ifCLASSINFOpdf
\else
\fi
\usepackage{mathrsfs}
\usepackage{graphicx}
\usepackage{epsfig}
\usepackage{amsmath}
\usepackage{bm}
\usepackage{float}
\usepackage{hyperref}
\usepackage{multirow}
\usepackage{color}
\usepackage{subfig}
\usepackage{amsmath}
\usepackage{algorithm}
\usepackage{algorithmic}
\usepackage{cite}
\usepackage{amsthm}
\usepackage{amssymb}
\usepackage{tabularx}
\usepackage{booktabs}
\usepackage{color}
\usepackage{float}

\usepackage{CJK}
\usepackage{amsmath}
\usepackage{stfloats, balance, epsfig}

\newenvironment{sequation*}{\begin{equation*}\small}{\end{equation*}}

\newtheorem{theorem}{Theorem}

\theoremstyle{definition}

\theoremstyle{remark}

\renewcommand{\algorithmicensure}{\textbf{Iteration:}}

\theoremstyle{proposition}

\long\def\symbolfootnote[#1]#2{\begingroup%
\def\thefootnote{\fnsymbol{footnote}}\footnote[#1]{#2}\endgroup}

\ifCLASSOPTIONcompsoc
\usepackage[nocompress]{cite}
\else
\usepackage{cite}
\fi
\usepackage{setspace}

\begin{document}

\title{\huge{Resource Allocation for Augmented Reality Empowered Vehicular Edge Metaverse}}
%\linespread{1.2}

\author{Jie Feng,
        Jun Zhao 
%\thanks{This work is surported by the National Key Research and Development Program of China (No.2020YFB1807500), the National Natural Science Foundation of China (No. 62102297, No. 62001357), Key Research and Development Program of Shaanxi (No. 2022GY-437), the Guangdong Basic and Applied Basic Research Foundation (No. 2020A1515110496, No. 2020A1515110079), the China Postdoctoral Science Foundation (No. 2021M692501), the Fundamental Research Funds for the Central Universities (No. XJS210105, No. XJS210107), and the Open Project of Shaanxi Key Laboratory of Information Communication Network and Security (No. ICNS202005). (\textit{Corresponding authors}: \textit{Qingqi Pei} and \textit{Wenjing Zhang})}
{\thanks{The authors are with School of Computer Science and Engineering, Nanyang Technological University, Singapore. Jie Feng is also with State Key Laboratory of ISN, School of Telecomm. Engineering, Xidian University, Xi'an, Shaanxi, China. Emails: jiefengcl@163.com, junzhao@ntu.edu.sg. }
%\thanks{Jinsong Wu is with the School of Artificial Intelligence, Guilin University of Electronic Technology, Guilin, China, and Department of Electrical Engineering, Universidad de Chile, Santiago, Chile (e-mail: wujs@ieee.org).}
}
}

%\thanks{Copyright (c) 2015 IEEE. Personal use of this material is permitted. However, permission to use this material for any other purposes must be obtained from the IEEE by sending a request to pubs-permissions@ieee.org.}
%\thanks{This work is jointly supported by Project 61271182 and 61302080 of the National Natural Science Foundation of China.
%}
%\thanks{L. Chen, H. Ji and G. Liu are with the Key Laboratory of Universal Wireless Communications, Ministry of Education, Beijing University of Posts and Telecommunications,
%Beijing, P.R. China (e-mail:chenlei1989bupt@gmail.com, jihong@bupt.edu.cn and buptgangliu@gmail.com).}
%\thanks{F. R. Yu is with the Dept. of Systems and Computer Eng., Carleton University, Ottawa, ON, Canada (e-mail: Richard.yu@carleton.ca).}
%\thanks{V. C. M. Leung is with the Department of Electrical and Computer Engineering, the University of British Columbia, Vancouver, BC V6T 1Z4 Canada
%(e-mail: vleung@ece.ubc.ca).}}
\maketitle
\vspace{-2.2cm}

\begin{abstract}
Metaverse is considered to be the evolution of the next-generation networks, providing users with experience sharing at the intersection between physical and digital. Augmented reality (AR) is one of the primary supporting technologies in the Metaverse, which can seamlessly integrate real-world information with virtual world information to provide users with an immersive interactive experience. Extraordinarily, AR has brought new opportunities for assisting safe driving. Nevertheless, achieving efficient execution of AR tasks and increasing system revenue are the main challenges faced by the Metaverse's AR in-vehicle applications. To address these challenges, in this paper, we are the first to propose an efficient resource allocation framework for AR-empowered vehicular edge Metaverse to improve system utility. For this purpose, we formulate an optimization problem featuring multidimensional control to concurrently maximize data utility at the Metaverse operator side and minimize energy consumption at the vehicles' side, which jointly considers the computational resource allocation on the Metaverse server, and AR vehicles' CPU frequency, transmit power, and computation model size. Notwithstanding, the major impediment is how to design an efficient algorithm to obtain the solutions of the optimization. Wherefore, we do this by decoupling the optimization variables. We first derive the optimal computation model size by the binary search, followed by obtaining the optimal power allocation by the bisection method and finding a closed-form solution to the optimal CPU frequency of AR vehicles, and finally,  attain the optimal allocation of computational resource on the server by the Lagrangian dual method. To estimate the performance of our proposed scheme, we establish three baseline schemes as a comparison, and simulation results manifest that our proposed scheme can balance the operator's reward and the energy consumption of vehicles. \vspace{-10pt}

\end{abstract}
\begin{IEEEkeywords}
~\vspace{-30pt} \newline Metaverse, wireless communications, resource allocation, optimization.
\end{IEEEkeywords}

\section{Introduction}\label{sec:1}
The Metaverse is considered as a disruptive digital revolution after the information and web Internet, known as the evolution of next-generation networks, which will completely transform the way we live and work \cite{ning2021survey}. In the Metaverse, users can obtain a real-time immersive experience based on augmented reality (AR) and virtual reality (VR) technologies that tightly blend physical and digital reality. Meanwhile, since the Metaverse is a collective virtual open space, each user can contribute to the Metaverse by producing content and editing the world, creating a virtual world for interaction, connection, sharing, and collaboration \cite{cai2022compute}. As a consequence, users in the Metaverse can not only enjoy the services it provides but also decorate it, where users can obtain corresponding benefits through this interaction. As the best medium of the Metaverse, games naturally have virtual scenes and virtual avatars, and have social, economic, and other attributes. For example, Axie Infinity is a play-to-earn game based on the concept of the metaverse. In this game, any player can get token bonuses by contributing to the ecosystem while enjoying the joy of the game \cite{dowling2022non}.

Particularly in the Internet of Vehicles (IoV), the visibility of objects on the road (such as pedestrians, surrounding vehicles, and stop signs) may be limited due to extreme weather, occlusion, and lighting conditions, which brings great safety concerns to traffic. By employing AR technology in vehicles, it can identify objects in the real environment and share the identified information with other vehicles in the Metaverse, increasing the safety of driving \cite{zhou2019enhanced,braud2022scaling}. Object detection is usually developed based on intelligent machine learning frameworks \cite{liu2019edge}. Nevertheless, as the popularity of sophisticated services accelerates, vehicles may have insufficient on-board computing capability \cite{9686591}. Vehicular edge computing (VEC) is an emerging architecture that provides a foundation for in-vehicle applications to improve service availability by leveraging the concept of computation offloading \cite{sonmez2020machine}. VEC improves user experience by offloading computation to an   edge infrastructure such as road side units (RSUs), base stations (BS), and access points. Hence, it can handle complex operations that cannot tolerate delays.

Edge computing is known as one of the supporting technologies of the Metaverse and is also becoming its main infrastructure \cite{dhelim2022edge}. In VEC, the Metaverse services are deployed in the VEC servers to provide AR applications for vehicles, forming  AR-enabled vehicle edge Metaverse.
On the one hand, to alleviate the computational overhead, vehicles only transmit the collected image/video frame of the roadside to the Metaverse system to expand the element library and obtain corresponding compensation. Accordingly, the vehicles will consume a part of the energy, including the image conversion energy consumption and transmission energy consumption.

On the other hand, the VEC servers perform object detection on the collected information. When vehicles need these results, they fetch them from the Metaverse and render them to the screens (e.g., windshields) and pay the related fee. From the perspective of the Metaverse operator, the more information the vehicles contribute, the more rewards will be obtained, but the overhead of the vehicles will increase. Therefore, how to balance the energy consumption of vehicles and the operator's reward is a problem that needs to be solved.

In this paper, we investigate resource allocation  in AR-empowered vehicular edge Metaverse. Specifically, we formulate an optimization to maximize the Metaverse operator's reward by jointly optimizing AR vehicles' CPU frequency, transmit power and computation model size, and the computational resource allocation on the Metaverse server. The main contributions of this paper are outlined as follows:
\begin{enumerate}
  \item We propose an efficient resource allocation framework for AR-empowered vehicular edge Metaverse to investigate the performance of the system. In this framework, the Metaverse focuses on AR applications for safe driving. With the AR technique, vehicles can obtain the real scene along the way in real-time.
  \item We formulate an optimization problem to simultaneously achieve the operator's reward maximization and the vehicles' energy consumption minimization.
  \item With the variables highly coupled, the formulated problem is difficult to solve directly. Hence, we decouple the optimization variables to develop an efficient algorithm. Specifically, the optimal solution of the computation model size is derived by the binary search. Moreover, the optimal power allocation is obtained by the bisection method. In addition, the optimal CPU frequency of AR vehicles is attained by a closed-form solution. Finally, the optimal allocation of computational resource  on the server is obtained by the Lagrangian dual technique.
  \item Our simulation  exhibits that the proposed algorithm performs well with regard to the operator's reward and energy consumption, and the   algorithm converges quickly. Furthermore, our scheme balances the operator's reward and the energy consumption of vehicles.

\end{enumerate}

The rest of this paper is structured as follows. In Section \ref{sec:11}, we summarize related works on Metaverse. Section \ref{sec:2} shows the system model and problem formulation. Section \ref{sec:3} proposes the design of the algorithms for CPU frequency of AR vehicle, transmit power for communication model, computation model size of AR vehicle, and computational resource allocation on the Metaverse server. Section \ref{sec:41} analyzes the complexity of the proposed algorithms. Then, the evaluation of the system performance in shown in Section \ref{sec:4}. Eventually, in Section \ref{sec:5}, we summarize the proposed scheme.
\section{Related Works}\label{sec:11}
Recently, Metaverse has received unprecedented attention from academia and industry. In this section, we first review the current research on the Metaverse and then exhibit the differences between the current works and our work.

Considering the applications of Metaverse to be data-intensive, Cai \textit{et al.} \cite{cai2022joint} designed an optimal control strategy to achieve efficient and real-time delivery of Metaverse applications by jointly orchestrating computing, caching, and communication resources. In \cite{duan2021metaverse}, Duan \textit{et al.} highlighted the application of Metaverse in the social good and presented a three-layer architecture for the Metaverse from a macroscopic point of view, including infrastructure, interaction, and ecosystem. Furthermore, Ning \textit{et al.} \cite{ning2021survey} examined the Metaverse from five perspectives, which refer to network infrastructure, VR object connection and convergence, general-purpose technology, and management technology. Given the recent advances in Metaverse, Huynh-The \textit{et al.} \cite{huynh2022artificial} explored the role of artificial intelligence in the development of the Metaverse, with particular emphasis on the application of machine learning and deep learning in the Metaverse. Except for \cite{cai2022joint}, the above works \cite{duan2021metaverse,ning2021survey,huynh2022artificial}
investigate the Metaverse from an overview rather than a technical perspective.
As we all know, AR technology occupies an important position in the Metaverse, so it is very important to study the application of AR in Metaverse \cite{siriwardhana2021survey}.

Thus, Liu \textit{et al.} \cite{liu2018dare} pointed out that edge computing can provide reliable AR services for Metaverse users, and formulated an optimization problem to simultaneously minimize the service latency and maximize the quality of augmentation. Since the majority of Metaverse applications are computationally intensive, Joseph \textit{et al.} \cite{redmon2017yolo9000} mainly studied how to efficiently allocate computing resources to process the environmental information uploaded by users from the perspective of computing resources, where the system performed the object detection task. In \cite{wang2020user}, Wang \textit{et al.} proposed a user preference-based energy-aware AR system, with the goal of minimizing the per-frame energy consumption of users by dynamically adjusting bandwidth allocation, CPU frequency, and data size in accordance with user preferences. However, the existing works \cite{wang2020user,liu2018dare,redmon2017yolo9000}
mainly study the AR digital experience in the Metaverse from the perspective of users without considering the operator's reward.
\section{System Model and Problem Formulation}\label{sec:2}
In this section, we first present the system scenario. Afterwards, we explain the local image conversion and communication model as well as the optimization objective in turn, respectively. Eventually, an optimization problem is formulated to maximize the system utility.
\subsection{System Scenario}\label{S21}
We consider an AR-empowered vehicular edge Metaverse system with an RSU and $N$ AR vehicles, signified by the set $\mathcal{N}=\{1, 2, \ldots, N\}$, where a VEC server is installed on the RSU. Metaverse services are deployed in the VEC server, wherefore vehicles can straightforwardly access its services through wireless communication \cite{ning2021survey}. As shown in Fig. 1, in the Metaverse, vehicles can have a comprehensive understanding of different objects and instances in the real world through AR technology. Furthermore, the vehicles can map real-world scenarios processed along the way onto the windshield via an AR-Head-up Display (AR-HUD) to enhance the driving experience. We assume that the AR-empowered vehicular edge Metaverse focuses on AR applications for object detection \cite{wang2020user}. The vehicle transmits the collected image/video frames to the Metaverse system. The object detection is performed on the Metaverse server (i.e., the VEC server), and the result is returned to the vehicles that access the Metaverse service. The results  are rendered on the vehicle screen. As specified in \cite{du2021optimal}, we assume that the vehicle does not move out of the RSU coverage area during the period. Simultaneously, the speed and the direction of each vehicle remain unchanged during that time \cite{bozorgchenani2021computation}.

In the scenario, there are contributing vehicles, a VEC server, and requesting vehicles, which are described as follows:

\begin{itemize}
  \item \textit{Contributing vehicles}: Contributing vehicles submit captured environmental information to Metaverse to enrich its scenes, making them the true creators of Metaverse, while they get rewarded accordingly. Meanwhile, they can also enjoy other Metaverse services for a more immersive experience.
  \item \textit{VEC server}: The VEC server not only deploys the Metaverse service but also provides computing services for users. In this paper, the VEC server can provide object detection applications to reduce the energy consumption of local vehicles. In the virtual open metaverse, the detection results can be provided to the requesting vehicles, thereby promoting the development of this ecosystem.
  \item \textit{Requesting vehicles}: Requesting vehicles are the vehicles requiring the Metaverse service. They can obtain results from the Metaverse service provider (i.e., the VEC server) and render images locally.
\end{itemize}

\begin{figure*}[t]
\centerline{\includegraphics[width=7.2in,height=3.0in]{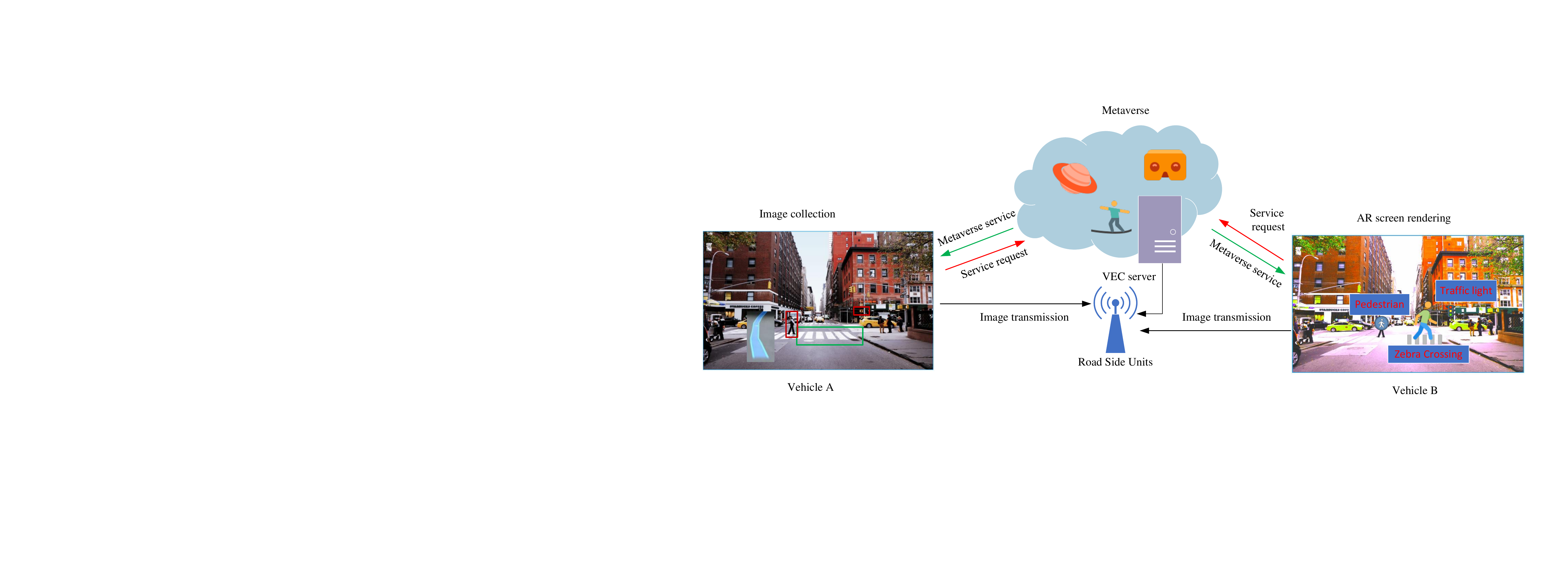}}
\caption{The system framework. }\label{fig:SEf1}
\end{figure*}

For the Metaverse service, we can consider that the contributing vehicles and VEC server form the set of ``producers'', and the requesting vehicles are the ``consumers''. In Figure~\ref{fig:SEf1}, each vehicle can be a contributing vehicle at the time when it is contributing, or can be a requesting vehicle when it is requesting. In this paper, we  focus on how the Metaverse service is constructed by the ``producers''; i.e., the part where the contributing vehicles and VEC server together build the Metaverse service. Involving requesting vehicles in the research can be a future direction.

% the generation stage of results\footnote{The results here refer to the value that users create for the Metaverse. 
%   %Users and vehicles represent objects of the Metaverse service, that is, they are equivalent.
%   }, and the screen rendering is beyond the scope of the paper.

\subsection{Local Image Conversion Model}\label{S22}
Before image transmission, vehicles need to do some pre-processing tasks such as camera sampling and image conversion. Since the images directly collected by cameras are in YUV format, it is necessary to convert YUV to RGB for later display \cite{yang2007fast}. The image conversion is processed via the AR vehicle's CPU. The technology of Dynamic Voltage and Frequency Scaling (DVFS)  is designed to save power by adjusting supply voltage and clock frequency \cite{eyerman2011fine}. The CPU power consumption for image conversion is modeled as $P_n^{cv}=P_n^{d}+P_n^{s}$, where $P_n^{d}$ and $P_n^{s}$ are dynamic and static power consumption, respectively. The superscripts ``$^{cv}$'', ``$^{d}$'', and ``$^{s}$'' in $P_n^{cv}$, $P_n^{d}$, and $P_n^{s}$ represent conversion, dynamic, and static, respectively. In this paper, $P_n^{s}$ is set to be a constant $\zeta_n$.
The dynamic power consumption for image conversion of AR vehicle $n$ is  modeled as $P_n^{d}\propto V_n^2f_n$, where $V_n$ and $f_n$ denote the supply voltage and frequency for the CPU of AR vehicle $n$, respectively. Since each $f_n$ needs to match with a specific $V_n$, the relationship between the frequency and the supply voltage is described as $V_n\propto f_n$ \cite{eyerman2011fine}. Consequently, the per frame energy consumption for the image conversion of AR vehicle $n$ can be expressed as
\begin{eqnarray}
E_n^{cv}=(\kappa_n f_n^3+\zeta_n)T_{n}^{cv},
\end{eqnarray}
where $\kappa_n$ denotes the effective switched capacitance of the CPU and $T_{n}^{cv}$ denotes the per frame computing time for the image conversion of AR vehicle $n$. Experiments show that the per frame computing latency mainly depends on the CPU frequency, and its value does not change much when the frame resolution increases \cite{wang2019energy}. 
Therefore, $T_{n}^{cv}$ can be given by $T_{n}^{cv}=\frac{C_n}{f_n},$
% \begin{eqnarray}
% T_{n}^{cv}=\frac{C_n}{f_n},
% \end{eqnarray}
where $C_n$ denotes the computation workload (in CPU cycles).
\subsection{Communication Model}\label{S23}
Similar to \cite{zeng2020volunteer,wu2020fog}, the well known orthogonal frequency division multiplexing (OFDM) protocol can be used for the channel model of vehicular communication. Let $P_n$ and $h_n$ be the transmit power and channel gain of AR vehicle $n$, respectively. Thus, the transmit rate of AR vehicle $n$ is given by
\begin{eqnarray}
R_n=B_n\log_2(1+\frac{P_nh_n}{B_n\sigma^2}),
\end{eqnarray}
where $B_n$ is the bandwidth derived by AR vehicle $n$ and $\sigma^2$ denotes the noise power density.

Since the communication delay is jointly determined by the data rate and the data size of the image frame transmitted by AR vehicles, $s_n$ (pixels) in this paper represents the computation model size of AR vehicle $n$. In object detection, the VEC server will receive an image frame with the resolution of $s_n \times s_n$ from AR vehicle $n$ to attain the corresponding detection accuracy (we assume ``square images'' for the frames in this paper for ease of exposition). Then, the data size of an image frame is $\varphi s_n^2$, where $\varphi$ denotes the number of bits required for the information carried by a pixel.

Accordingly, the communication delay required for AR vehicle $n$ to transmit an image frame is expressed as $T_n^{com}=\frac{\varphi s_n^2}{R_n}.$
Then the energy consumption of AR vehicle $n$ is given by
\begin{eqnarray}
E_n^{com}=P_nT_n^{com}=\frac{P_n\varphi s_n^2}{R_n}.
\end{eqnarray}

\subsection{Optimization Objective}\label{S24}

We consider the optimization objective as the  system utility defined below, which can also referred to the social welfare of the system. The concept of social welfare originates from economics and has been adopted by wireless researchers~\cite{jiao2020toward}. As explained in the Section~\ref{S21} on the system scenario, we focus on  the part where the contributing vehicles and VEC server together build the Metaverse service. Then the network utility means the sum of the utilities of the contributing vehicles and VEC server for the construction of the Metaverse service. Note that utility equals the gross profit minus the cost. The gross profit part comes from the data sent from the contributing vehicles to the VEC server (i.e., the Metaverse server). The data is used to build the Metaverse service which can make profits; in a nutshell, data is profitable~\cite{singh2022interplay}. We assume that the $\varphi s_n^2$ bits of data sent from AR vehicle $n$ to the Metaverse server can induce a profit of $\rho_n\ln(1+\varphi s_n^2)$, where $\rho_n$ is a weight factor depending on the data type, the data quality, etc. Such profit of $\rho_n\ln(1+\varphi s_n^2)$ may be shared between the Metaverse server and AR vehicle $n$, but how the sharing is done is not relevant to our optimization objective since we care about the network utility in which the sum of the utilities do not depend on the sharing details. We choose the logarithmic function to model the profit because of its positive first-order derivative and negative second-order derivative. The positive first-order derivative (i.e., the monotonically increasing property) reflects that the higher the image resolution provided by AR vehicles, the more the utility of the data obtained by the Metaverse operator. The negative second-order derivative (i.e., concavity) means that the rate of increase is slowing. In other words, the marginal profit is decreasing; i.e., when there is already sufficient data at the Metaverse server, the additional profit from collecting new data may not be much. Using the logarithmic function to model the profit has also been adopted by a classical work~\cite{yang2012crowdsourcing} on crowdsourcing by Yang \textit{et~al.} Also, using the natural logarithm does not lose generality. The reason is that if we use a logarithm with base $b$, we can still  have the natural logarithm as the expression since the constant factor can be absorbed into the coefficient before the natural logarithm; i.e., $\rho_n\log_b(1+\varphi s_n^2)=\frac{\rho_n}{\ln b}\cdot\ln(1+\varphi s_n^2)$.  From the above discussion, the gross profit of the system is given by
\begin{eqnarray}
U(\boldsymbol s)=\sum\limits_{n\in\mathcal{N}}\rho_n\ln(1+\varphi s_n^2). \label{reward}
\end{eqnarray}
where $\boldsymbol s\triangleq [s_1, s_2, \ldots, s_N]$  (``$\triangleq$'' means ``equal to by definition''). Since we do not need to discuss how the gross profit is shared between the Metaverse server and AR vehicles. For ease of presentation, we can refer to Equation~(\ref{reward}) as the reward of the Metaverse operator.

For the cost part, we consider only energy consumption in this paper (incorporating latency can be a future direction). From Section~\ref{S22} above, the energy consumption of AR vehicle $n$ is $E_n^{cv}+E_n^{com}$. Below we compute the energy consumption of the Metaverse server.

% In this paper, we consider the Metaverse operator's reward as our performance metric, which is defined as the net data utility after removing the energy cost and the cost paid. Concretely, we give its mathematical definition according to \cite{feng2019computation,yang2012crowdsourcing}. The utility of $\varphi s_n^2$ bits data collected by the Metaverse server is expressed by the logarithmic function. We can observe that the utility function is monotonically increasing. The fact reflects that the higher the image resolution provided by AR vehicles, the more the utility of the data obtained by the Metaverse operator. 

After the VEC server receives the image frame of AR vehicles, it needs to complete the object detection. Thus, the time required for the server to infer one image frame of AR vehicle $n$ is 
\begin{eqnarray}
T_n^{ser}=\frac{\varphi s_n^2c_n}{f_n^s}, \label{eqTnser}
\end{eqnarray}
where $f_n^s$ (in cycle/s) denotes the computation resource on the VEC server allocated to process the image frame of AR vehicle $n$, and $c_n$ (in cycle/bit) denotes the number of CPU cycles required to process $1$ bit. Similar to the first author's prior work \cite{feng2021min}, we model the power consumption of the server as $\kappa_{ser}(f_n^s)^3$, where $\kappa_{ser}$ is the effective switched capacitance of the server CPU. Such cube (i.e., third-power) relation is also adopted in many other papers. Since energy equals power multiplied by time, the energy consumed for the server to infer one image frame of AR vehicle $n$ is given by   
\begin{eqnarray}
E_n^{ser}=\kappa_{ser}(f_n^s)^3T_n^{ser}=\kappa_{ser}(f_n^s)^2\varphi s_n^2c_n,
\end{eqnarray}
where the last step uses Equation~(\ref{eqTnser}).

Then, taking into account the gross profit and energy consumption cost discussed above, the system utility can be denoted by
\begin{eqnarray}
&&\!\!\!\!\!\!\!\!\!\!\!\!\!\!\!\!\!\!\!\!\!C(\boldsymbol f, \boldsymbol s,\boldsymbol P,\boldsymbol f^s)=U(\boldsymbol s)-\sum\limits_{n\in\mathcal{N}}\left(\beta_nE_n^{ser}+\gamma_n(E_n^{cv}+E_n^{com})\right),
\end{eqnarray}
where $\boldsymbol f\triangleq  [f_1, f_2, \ldots, f_N]$, $\boldsymbol s\triangleq [s_1, s_2, \ldots, s_N]$, $\boldsymbol P\triangleq  [P_1, P_2, \ldots, P_N]$, $\boldsymbol f^s\triangleq  [f^s_1, f^s_2, \ldots, f^s_N]$. We can understand $\beta_n$ and $\gamma_n$ as weights which makes our optimization problem flexible for various cases. Moreover, $\beta_n$ and $\gamma_n$ can also be interpreted as  the price of unit energy paid by the server and AR vehicle $n$, respectively (in this case, $\beta_n$ may be the same for all $n$, but we stil write $\beta_n$ for generality).

All the notations used are listed in Table I.
%\begin{table*}%table* ÊÇË«À¼Ä£Ê½ÏÂµÄ¿çÀ¸

\begin{table*}[!t] 
\caption{Notation used in this paper.}
\label{list1}
\setlength{\tabcolsep}{1pt} \hspace{-10pt}
\small
\begin{tabular}{|c|c|c|c|}
\hline
\multicolumn{1}{|c|}{\bf{Symbol}}     & \multicolumn{1}{c|}{\bf{Definitions }}       & \multicolumn{1}{c|}{\bf{Symbol}}  & \multicolumn{1}{c|}{\bf{Definitions }} \\ \hline
\multicolumn{1}{|c|}{$\mathcal{N}$ (resp.,$N$)}             & \multicolumn{1}{c|}{Set (resp., Number) of AR vehicles}    & \multicolumn{1}{c|}{$T$}    & \multicolumn{1}{c|}{Length of the fixed time window}    \\ \hline
\multicolumn{1}{|c|}{$P_n^{cv}$}             & \multicolumn{1}{c|}{\shortstack{Per frame power consumption for image \\ conversion of AR vehicle $n$}}    & \multicolumn{1}{c|}{$E_n^{cv}$}  & \multicolumn{1}{c|}{\shortstack{Per frame energy consumption for the \\image conversion of AR vehicle $n$}} \\ \hline
\multicolumn{1}{|c|}{$V_n$}             & \multicolumn{1}{c|}{Supply voltage of AR vehicle $n$ CPU}    & \multicolumn{1}{c|}{$f_n$}  & \multicolumn{1}{c|}{CPU frequency of AR vehicle $n$} \\ \hline
\multicolumn{1}{|c|}{$\alpha_1,\alpha_2$ }       & \multicolumn{1}{c|}{\shortstack{Positive coefficients for energy \\ consumption in AR vehicle}}   & \multicolumn{1}{c|}{$a,b,c,d$}  & \multicolumn{1}{c|}{\shortstack{Constant coefficients for computing \\ time in AR vehicle}} \\ \hline
\multicolumn{1}{|c|}{$T_n^{cv}$ }       & \multicolumn{1}{c|}{\shortstack{Computing time for converting a frame \\ image in AR vehicle $n$}}   & \multicolumn{1}{c|}{$R_{n}$}  & \multicolumn{1}{c|}{Transmit rate of AR vehicle $n$} \\ \hline
\multicolumn{1}{|c|}{$B_n$ }       & \multicolumn{1}{c|}{Bandwidth derived by AR vehicle $n$}   & \multicolumn{1}{c|}{$\sigma$} & \multicolumn{1}{c|}{Noise power in communication model} \\ \hline
\multicolumn{1}{|c|}{$h_n$ }       & \multicolumn{1}{c|}{Channel gain of AR vehicle $n$}   & \multicolumn{1}{c|}{$s_n$} & \multicolumn{1}{c|}{Computation model size of AR vehicle $n$ } \\ \hline
\multicolumn{1}{|c|}{$\varphi$}       & \multicolumn{1}{c|}{\shortstack{Number of bits required for the information \\ carried by a pixel}}   & \multicolumn{1}{c|}{$T_n^{com}$}  & \multicolumn{1}{c|}{\shortstack{Communication delay required for AR vehicle \\ to transmit an image frame $n$}}\\ \hline
\multicolumn{1}{|c|}{$E_n^{com}$ }       & \multicolumn{1}{c|}{Energy consumption of AR vehicle $n$}   & \multicolumn{1}{c|}{$\rho_n$} & \multicolumn{1}{c|}{Weight factor building on the type of data} \\ \hline
\multicolumn{1}{|c|}{$U(\boldsymbol s)$ }   &    \multicolumn{1}{c|}{Sum data utility of the Metaverse operator}   & \multicolumn{1}{c|}{$T_n^{ser}$}   & \multicolumn{1}{c|}{\shortstack{Time required for the server to infer an \\ image frame of AR vehicle $n$}} \\ \hline
\multicolumn{1}{|c|}{$f_n^s$ }    &   \multicolumn{1}{c|} {\shortstack{computational resource allocated \\ on the Metaverse server  to \\ process data from AR vehicle $n$}}   & \multicolumn{1}{c|}{$c_n$} & \multicolumn{1}{c|}{\shortstack{Number of CPU cycles required to \\ process one bit}} \\ \hline
\multicolumn{1}{|c|}{$E_n^{ser}$}       & \multicolumn{1}{c|}{Energy consumed of the Metaverse server}   & \multicolumn{1}{c|}{$\kappa_n$, $\kappa_{ser}$} & \multicolumn{1}{c|}{\shortstack{Effective switched capacitance of \\  AR vehicles or server CPU processor}} \\ \hline
\multicolumn{1}{|c|}{$s_{min},s_{max}$}       & \multicolumn{1}{c|}{\shortstack{Minimum and maximum value \\ of computation model size of any AR vehicle}}   & \multicolumn{1}{c|}{$f_{min},f_{max}$} & \multicolumn{1}{c|}{\shortstack{Minimum and maximum CPU \\ frequency of AR vehicle}} \\ \hline
\multicolumn{1}{|c|}{$F$}       & \multicolumn{1}{c|}{\shortstack{Total computing capacity of the \\ Metaverse server}}   & \multicolumn{1}{c|}{$\delta_n$} & \multicolumn{1}{c|}{\shortstack{Minimum analytics accuracy requirement \\ of AR vehicle $n$}} \\ \hline
\multicolumn{1}{|c|}{$C_n$}       & \multicolumn{1}{c|}{Computation workload of AR vehicle $n$}   & \multicolumn{1}{c|}{$\nu,\boldsymbol \mu$} & \multicolumn{1}{c|}{Lagrange multipliers} \\ \hline
\end{tabular}
\end{table*}
\subsection{Problem Formulation}\label{S25}
The concrete problem in this paper is to jointly optimize the CPU frequency  $\boldsymbol f$ of AR vehicle, the transmit power $\boldsymbol P$ for communication model, the computation model size  $\boldsymbol s$ of AR vehicle, and the computational resource $\boldsymbol f^s$ allocated on the Metaverse server to handle information from AR vehicles.
We aim to maximize the system utility when the Metaverse server obtains per frame image from each AR contributing vehicle. Mathematically, the proposed optimization problem can be formulated as follows (the constraints' meanings are deferred to the next paragraph):
\begin{eqnarray}\label{P1}
\text{Problem (P1):~~~~~} &&\!\!\!\!\!\!\!\!\!\!\!\!\!\!\!\!\!\max\limits_{\boldsymbol f, \boldsymbol P, \boldsymbol s, \boldsymbol f^s} C(\boldsymbol f, \boldsymbol s,\boldsymbol P,\boldsymbol f^s)\nonumber\\
&&\!\!\!\!\!\!\!\!\!\!\!\!\mathrm{s.t.}~(\mathrm{C1}):~s_n\in\{s_{min},\ldots,s_{max}\},~~\forall n, \nonumber\\
&&\!\!\mathrm{(C2)}:~f_{min}\leq f_n\leq f_{max},~~\forall n, \nonumber \\
&&\!\!\mathrm{(C3)}:~\sum\limits_{n\in\mathcal{N}}f_n^s\leq F, \nonumber\\
&&\!\!\mathrm{(C4)}:~0< P_n\leq P_n^{max},~~\forall n,\\
&&\!\!\mathrm{(C5)}:~\varepsilon(s_n^2)\geq\delta_n,~~\forall n,\nonumber\\
&&\!\!\mathrm{(C6)}:~\frac{\varphi s_n^2}{R_n}+\frac{\varphi s_n^2 c_n}{f_n^s}+T_n^{cv}\leq T,~~\forall n. \nonumber\\
&&\!\!\mathrm{(C7)}:~f_n^s>0,~~\forall n. \nonumber
\end{eqnarray}

In Problem (P1) above, $\mathrm{(C1)}$ is the constraint of computation model size $s_n$, where $s_{min}$ and $s_{max}$ denote the minimum and maximum value of computation model size of any AR vehicle, respectively\footnote{When there is a need to express all the discrete values of Constraint $\mathrm{(C1)}$, we assume there are $M$ values and write $\{s_{min},\ldots,s_{max}\}$ as the following sorted set (in ascending order): $\{s^{(1)},s^{(2)},\ldots,s^{(M)}\}$, where $s_{min}=s^{(1)}$ and $s_{max}=s^{(M)}$. \label{discrete}}. From Constraint $\mathrm{(C1)}$ and Footnote~\ref{discrete} below, $s_n$ is a discrete variable whose value relies on the computational models available in the AR-empowered vehicular edge Metaverse system.  $\mathrm{(C2)}$ means the CPU frequency constraint of AR vehicle $n$, where $f_{min}$ and $f_{max}$ are respectively the minimum and maximum CPU frequency of AR vehicle. $\mathrm{(C3)}$ implies that the sum of computational resources allocated to AR vehicles cannot exceed the total computing capacity (i.e., the maximum CPU frequency) of the Metaverse server, denoted by $F$. In $\mathrm{(C4)}$, $P_n^{max}$ is the maximum transmit power allowed by AR vehicle $n$. This constraint ensures that the transmit power configured to AR vehicles is within the acceptable range. In order to ensure that the Metaverse system can provide a better quality of service, the object recognition accuracy cannot be too low, so $\mathrm{(C5)}$ represents the constraint of the analytic accuracy, where $\delta_n$ is the minimum analytics accuracy requirement of AR vehicle $n$. In \cite{liu2018edge}, the analytics accuracy is modeled as $\varepsilon(s_n^2)=1-1.578e^{-6.5\times 10^{-3}s_n}$. $\mathrm{(C6)}$ denotes that the service latency of the Metaverse operator is no larger than a fixed value $T$.

\section{Design of Resources Allocation}\label{sec:3}
In this section, we develop algorithms to resolve the optimization problem in (\ref{P1}) separately for the CPU frequency of AR vehicles, the transmit power for the communication model, the computation model size of AR vehicles, and the computational resource allocation on the Metaverse server. 

As discussed previously, since problem (\ref{P1}) involves both continuous variables ($\boldsymbol P$,$\boldsymbol f$,$\boldsymbol f^s$) and discrete variable $\boldsymbol s$, it is a mixed-integer non-linear programming problem, which is difficult to tackle directly \cite{belotti2013mixed}. Therefore, to solve the problem, we relax $s_n$ to $[s_{min}, s_{max}]$. Then, the problem after relaxation becomes
\begin{eqnarray}
&&\!\!\!\!\!\!\!\!\!\!\!\!\!\!\!\!\!\max\limits_{\boldsymbol f, \boldsymbol P, \boldsymbol s, \boldsymbol f^s} C(\boldsymbol f, \boldsymbol s,\boldsymbol P,\boldsymbol f^s)\nonumber\\ &&\!\!\!\!\!\!\!\!\!\!\!\!\mathrm{s.t.}~~(\mathrm{C1})':~s_{min}\leq s_n\leq s_{max},~~\forall n, \nonumber\\[-10pt]
&&(\mathrm{C2}), (\mathrm{C3}),(\mathrm{C4}), (\mathrm{C5}), (\mathrm{C6}),(\mathrm{C7}).\nonumber\\[-30pt]
\end{eqnarray}
\subsection{Computation Model Size Optimization}\label{S31}
 For a given CPU frequency of AR vehicle $\boldsymbol f$, transmit power for communication model $\boldsymbol P$, and computational resource allocation on the Metaverse server, the optimal solutions of computation model size can be obtained by solving the following problem.
\begin{eqnarray}\label{P2}
\text{Problem (P2):~~~~~}&&\!\!\!\!\!\max\limits_{\boldsymbol s} \sum\limits_{n\in\mathcal{N}}\left(\rho_n\ln(1+\varphi s_n^2)-\beta_n\kappa_{ser}(f_n^s)^2\varphi s_n^2c_n-\gamma_n\frac{P_n\varphi s_n^2}{R_n}\right) - \sum\limits_{n\in\mathcal{N}}\gamma_nE_n^{cv}\nonumber\\[-2pt]
&&\!\!\!\!\!\!\mathrm{s.t.}~(\mathrm{C1})':~s_{min}\leq s_n\leq s_{max},~~\forall n, \nonumber\\[-2pt]
&&\!\!~~~\mathrm{(C5)}:~s_n\geq\frac{1}{6.5\times10^{-3}}(\ln1.578-\ln(1-\delta_n)),~~\forall n,\nonumber\\[-2pt]
&&\!\!~~~\mathrm{(C6)}:~s_n\leq\sqrt{\frac{(T-T_{n}^{cv})R_nf_n^s}{\varphi f_n^s+\varphi c_nR_n}},~~\forall n,
\end{eqnarray}
where $\sum\limits_{n\in\mathcal{N}}\gamma_nE_n^{cv}$ in the objective function of Problem (P2) can be further removed since it does not depend on $\boldsymbol s$. 

% Since $s_n\geq0$, the constraint $\mathrm{(C6)}$ can be transformed into
% \begin{eqnarray}
% s_n\leq\sqrt{\frac{(T-T_{n}^{cv})R_nf_n^s}{\varphi f_n^s+\varphi c_nR_n}}.
% \end{eqnarray}

Since both the objective function to be maximized in Problem (P2) is concave and the constraints of Problem (P2) are linear, Problem (P2) is a convex optimization problem. Furthermore, we find that the constraints of Problem (P2) are for individual $s_n$, so its objective function can be dissolved into individual users. Wherefore, the solution of Problem (P2) is obtained \textbullet~either at some boundary point (i.e., a lower/upper bound of $s_n$ in the constraints of Problem (P2)) or \textbullet~at the maximum point of objective function of Problem (P2); i.e., at the maximum point of the following function $\text{Obj}_n(s_n)$:
\begin{align}
\text{Obj}_n(s_n) \triangleq   \rho_n\ln(1+\varphi s_n^2)-\beta_n\kappa_{ser}(f_n^s)^2\varphi s_n^2c_n-\gamma_n\frac{P_n\varphi s_n^2}{R_n}. \label{Objsn}
\end{align}
Specifically, the optimal computation model size includes the following cases, where $\varepsilon_1$ denotes the lower bound of $s_n$ in Constraint (C5) of Problem (P2), and $\varepsilon_2$ denotes the upper bound of $s_n$ in Constraint (C6) of Problem (P2); i.e.,  $\varepsilon_1\triangleq\frac{1}{6.5\times10^3}(\ln1.578-\ln(1-\delta_n))$ and $\varepsilon_2\triangleq\sqrt{\frac{(T-T_{n}^{cv})R_nf_n^s}{\varphi f_n^s+\varphi c_nR_n}}$.

\textit{Case 1:} $\varepsilon_1\geq\varepsilon_2$ and $\varepsilon_2\leq s_{min}$ and $\varepsilon_1\geq s_{max}$. Then Problem (P2) has no feasible solution.

\textit{Case 2:} $\varepsilon_1\geq s_{min}$ and $\varepsilon_2\geq s_{max}$. In this case, we can deduce that the value range of $s_n$ is in $[\varepsilon_1, s_{max}]$. Then, the solution $s_n$ to Problem (P2)  is given by
\begin{eqnarray}\label{s1}
s_n=\left\{\begin{array}{ll}
\mathfrak{A}_n, &\textrm{if $\varepsilon_1\leq\mathfrak{A}_n\leq s_{max}$,}\\[-2pt]
\varepsilon_1, &\textrm{if $\mathfrak{A}_n\leq \varepsilon_1$,}\\[-2pt]
s_{max},&\textrm{if $\mathfrak{A}_n\geq s_{max}$,}\\[-2pt]
\end{array}\right.
\end{eqnarray}
where $\mathfrak{A}_n$ is defined as follows:
\begin{eqnarray}
\mathfrak{A}_n\triangleq\sqrt{\left[\frac{\rho_n}{\varphi\left(\beta_n\kappa_{ser}(f_n^s)^2c_n+\frac{\gamma_nP_n}{R_n}\right)\ln2}-\frac{1}{\varphi}\right]^{+}}, \label{eqmathfrakA}
\end{eqnarray}
where $[x]^+\triangleq \max\{0,x\}$. Note that we will also use $\mathfrak{A}_n$ of Equation~(\ref{eqmathfrakA}) in other cases below.

\textit{Case 3:} $\varepsilon_1\leq s_{min}$ and $\varepsilon_2\leq s_{max}$. In this case, we can deduce that the value range of $s_n$ is in $[s_{min},\varepsilon_2]$. Then, the solution $s_n$ to Problem (P2) is given by
\begin{eqnarray}\label{s2}
s_n=\left\{\begin{array}{ll}
\mathfrak{A}_n, &\textrm{if $s_{min}\leq\mathfrak{A}_n\leq \varepsilon_1$,}\\[-2pt]
s_{min}, &\textrm{if $\mathfrak{A}_n\leq s_{min}$,}\\[-2pt]
\varepsilon_2,&\textrm{if $\mathfrak{A}_n\geq\varepsilon_2$.} 
\end{array}\right.
\end{eqnarray}

\textit{Case 4:} $\varepsilon_1\geq s_{min}$ and $\varepsilon_2\leq s_{max}$. In this case, we can deduce that the value range of $s_n$ is in $[\varepsilon_1, \varepsilon_2]$. Then, the solution $s_n$ to Problem (P2) is given by
\begin{eqnarray}\label{s3}
s_n=\left\{\begin{array}{ll}
\mathfrak{A}_n, &\textrm{if $\varepsilon_1\leq\mathfrak{A}_n\leq \varepsilon_2$,}\\[-2pt]
\varepsilon_1, &\textrm{if $\mathfrak{A}_n\leq \varepsilon_1$,}\\[-2pt]
\varepsilon_2,&\textrm{if $\mathfrak{A}_n\geq \varepsilon_2$.}\\[-2pt]
\end{array}\right.
\end{eqnarray}

Since the computation model size $s_n$ in the original  Problem (P1) on Page~\pageref{P1} is a discrete variable, the result (denoted by 
$s_n^{\text{ContinuousOpt}}$)  obtained by one of (\ref{s1}), (\ref{s2}), and (\ref{s3}) (depending on which one is applicable) may not be the true solution to Problem (P1). The superscript ``ContinuousOpt'' means that $s_n^{\text{ContinuousOpt}}$ is obtained from Problem (P2) where $s_n$ is relaxed to be continuous. Hence, we develop a binary search algorithm to obtain the solution $s_n^{*}$ to Problem (P1), and the whole algorithm procedure is shown in Algorithm~1. Since there is a need to express all the discrete values of Constraint $\mathrm{(C1)}$ of Problem (P1), as noted in Footnote~\ref{discrete}, we assume there are $M$ values and write $\{s_{min},\ldots,s_{max}\}$ of Problem (P1) as the following sorted set (in ascending order): $\{s^{(1)},s^{(2)},\ldots,s^{(M)}\}$, where $s_{min}=s^{(1)}$ and $s_{max}=s^{(M)}$. 

\begin{algorithm}[!t]
\caption{Binary Search for Computation Model Size Allocation}
\label{alg:computation model size}
\begin{spacing}{1.2}\begin{algorithmic}[1]
\REQUIRE ~~\\
 $\bullet$ $s_n^{\text{ContinuousOpt}}$, which is obtained by one of (\ref{s1}), (\ref{s2}), and (\ref{s3}), depending on which one is applicable. \\
 $\bullet$  Set  $\{s^{(1)},s^{(2)},\ldots,s^{(M)}\}$.\\% and $\Omega=\emptyset$. 
 $\bullet$  Set $a=1$ and $b=M$.\\ %$K=M$ and $\Lambda=\{0,1,\ldots,M-1\}$.\\
\ENSURE ~~\\
\IF{$M=1$}
\STATE return $s^{(1)}$ as $s_n^{*}$ for $n$ from 1 to $N$.
\ENDIF 
\IF{$M=2$} 
\STATE return $s^{(1)}$ as $s_n^{*}$ if $\text{Obj}_n(s^{(1)}) \geq \text{Obj}_n(s^{(2)})$ and return $s^{(2)}$ as $s_n^{*}$ otherwise, for $n$ from 1 to $N$, where the function $\text{Obj}_n(\cdot)$ is defined in Equation~(\ref{Objsn}). 
\ENDIF
\IF{$M \geq 3$}
\FOR {$n$ from 1 to $N$}
\WHILE {$b-a>1$} 
\STATE $c=\lfloor\frac{a+b}{2}\rfloor$, where the floor function $\lfloor x \rfloor$ means the greatest integer as most $x$.\\
\IF{$s_n^{\text{ContinuousOpt}} > s^{(c)}$}
\STATE $a=c$.\\
%\STATE $\Omega=\Omega\cup\{[M/2]\}$, where $[x]$ means rounding.\\
%\STATE $\Lambda=\{[M/2],\ldots,M-1\}$, where $[x]$ means rounding.\\
\ELSIF{$s_n^{\text{ContinuousOpt}} < s^{(c)}$}
\STATE $b=c$.
%\STATE $\Omega=\Omega\cup\{[M/2]-1\}$, where $[x]$ means rounding.\\
%\STATE $\Lambda=\{0,1,\ldots[M/2]-1\}$, where $[x]$ means rounding.\\
\ELSE
\STATE return $s_n^{\text{ContinuousOpt}}$ as $s_n^{*}$.
\ENDIF 
\ENDWHILE
\STATE return $s^{(a)}$ as $s_n^{*}$ if $\text{Obj}_n(s^{(a)}) \geq \text{Obj}_n(s^{(b)})$ and return $s^{(b)}$ as $s_n^{*}$ otherwise. 
\ENDFOR
\ENDIF
\end{algorithmic}\end{spacing}
\end{algorithm}
\subsection{Transmit Power Allocation}
Given CPU frequency of AR vehicle $\boldsymbol f$, computation model size $\boldsymbol s$, and computational resource allocation on the Metaverse server, the transmit power allocation problem is recast as
\begin{eqnarray}\label{TP1}
&&\min\limits_{\boldsymbol P}\sum\limits_{n\in\mathcal{N}}\frac{\gamma_n\varphi s_n^2P_n}{R_n}\nonumber\\
&&\mathrm{s.t.}~\mathrm{(C4)}:~0< P_n\leq P_n^{max},~~\forall n,\\
&&~~~~~\mathrm{(C6)}:~\frac{\varphi s_n^2}{R_n}\leq T-(\frac{\varphi s_n^2 c_n}{f_n^s}+T_n^{cv}),~~\forall n,\nonumber
\end{eqnarray}

Then (\ref{TP1}) is equivalent to optimizing the energy consumption of individual AR vehicles; i.e, we can have an optimization problem for each $n$. In addition, due to $R_n=B_n\log_2(1+\frac{P_nh_n}{B_n\sigma^2})$, constraint (C6) is equivalent to 
\begin{align}
P_n \geq \mathcal{Q}_n \triangleq  \left( 2^{\frac{\mathcal{C}_n}{B_n}}-1\right) B_n \sigma^2 / {h_n} , \text{ where } \mathcal{C}_n\triangleq\frac{\varphi s_n^2f_n^s}{(T-T_n^{cv})f_n^s-c_n\varphi s_n^2}  . \label{ObjQn} 
\end{align}
Hence, the optimization problem for each $n$ can be given by
\begin{eqnarray}\label{TP2}
&&\min\limits_{P_n}F_n(P_n), \text{~for~} F_n(P_n) \triangleq \frac{\gamma_n\varphi s_n^2P_n}{R_n}= \frac{\gamma_n\varphi s_n^2P_n}{B_n\log_2(1+\frac{P_nh_n}{B_n\sigma^2})}, \\
&&\mathrm{s.t.}~\mathrm{(C8)}:~\mathcal{Q}_n \leq P_n\leq P_n^{max} . \nonumber 
\end{eqnarray}
If $\mathcal{Q}_n > P_n^{max} $, Problem (\ref{TP2}) has no solution. If $\mathcal{Q}_n \leq P_n^{max} $, we will use the following theorem to convert Problem (\ref{TP2}).

% To distinctly characterize the property of (\ref{TP2}), the following theorem is introduced.
\begin{theorem} \label{theorem1}
The objective function $F_n(P_n)$ of Problem
 (\ref{TP2}) is quasiconvex on $P_n$ since the nominator of $F_n(P_n)$ is convex and the denominator of $F_n(P_n)$ is concave.
Problem (\ref{TP2}) is quasiconvex on $P_n$.
\end{theorem}

\textit{Proof:} The theorem is proved in Appendix A.

Since $\frac{\text{nominator}}{\text{denominator}} \leq t$  is equivalent to $(\text{nominator} - t \times {\text{denominator}} \leq 0)$. To analyze the objective function of Problem
 (\ref{TP2}), we define a family of functions $\phi_n^{(t)}(P_n)$ as $(\text{nominator} - t \times {\text{denominator}})$; i.e., 
\begin{align}
\phi_n^{(t)}(P_n) \triangleq   \gamma_n\varphi s_n^2P_n-tB_n\log_2(1+\frac{P_nh_n}{B_n\sigma^2}). \label{ObjsnPn}
\end{align}
Then to solve Problem (\ref{TP2}), we will consider the following feasibility problem. 
\begin{eqnarray}\label{TP31}
&& \text{Find }P_n \\
&&\mathrm{s.t.}~\phi_n^{(t)}(P_n) \leq 0, \nonumber\\
&&~~~~~\mathrm{(C8)}:~\mathcal{Q}_n \leq P_n\leq P_n^{max} . \nonumber  
\end{eqnarray}

The relationships between Problem (\ref{TP2}) and Problem (\ref{TP31}) are as follows:
\begin{itemize}
\item When $t$ takes a certain value $\widetilde{t}$, if Problem (\ref{TP31}) is feasible (e.g., has a solution $\widetilde{P_n}$), then $\phi_n^{(\,\widetilde{t}\,)}(\widetilde{P_n}) \leq 0$ and $\mathcal{Q}_n \leq \widetilde{P_n}\leq P_n^{max}$. The inequality $\phi_n^{(\,\widetilde{t}\,)}(\widetilde{P_n}) \leq 0$ means $F_n(\widetilde{P_n}) \leq \widetilde{t}$. Hence, if Problem (\ref{TP31}) is feasible for $\widetilde{t}$, the optimal objective-function value of Problem (\ref{TP2}) is no greater than $\widetilde{t}$.
\item  When $t$ takes a certain value $\overline{t}$, if Problem (\ref{TP31}) is infeasible, then $\phi_n^{(\,\overline{t}\,)}(P_n) > 0$ for any $\mathcal{Q}_n \leq P_n\leq P_n^{max}$. The inequality $\phi_n^{(\,\overline{t}\,)}(P_n) > 0$ means $F_n(P_n) > \overline{t}$. Hence, if Problem (\ref{TP31}) is infeasible for $\overline{t}$, the optimal objective-function value of Problem (\ref{TP2}) is greater than $\overline{t}$. 
\end{itemize}

To solve the Problem (\ref{TP31}) on feasibility, we consider the following minimization problem. 
\begin{eqnarray}\label{TP33}
&& \min\limits_{P_n}\phi_n^{(t)}(P_n) \\ 
&&\mathrm{s.t.}~\mathrm{(C8)}:~\mathcal{Q}_n \leq P_n\leq P_n^{max} .\nonumber
\end{eqnarray}
From Theorem~\ref{theorem1}, $\phi_n^{(t)}(P_n)$ above is convex. We further have:
\begin{itemize}
\item 
If the optimal objective-function value of Problem (\ref{TP33}) is no greater than 0, then Problem (\ref{TP31}) is feasible and thus the corresponding $t$ provides an upper bound for the optimal objective-function value of Problem (\ref{TP2}).  
\item 
If the optimal objective-function value of Problem (\ref{TP33}) is  greater than 0, then Problem (\ref{TP31}) is infeasible and thus the corresponding $t$ provides a strict lower bound for the optimal objective-function value of Problem (\ref{TP2}).  
\end{itemize}

Based on the above, we can use the bisection method to solve Problem (\ref{TP2}). We start from a lower bound $0$ and an upper bound $F_n(P_n^{max}) $ for the optimal objective-function value of Problem (\ref{TP2}). Hence $[0,F_n(P_n^{max})] $ is the interval $[ t_{\text{lower}}, t_{\text{upper}}]  $ at the beginning. At each iteration, starting with the interval $[ t_{\text{lower}}, t_{\text{upper}}]  $, we compute the middle point of the interval $t_{\text{middle}}=( t_{\text{lower}} + t_{\text{upper}} )/2$. Then we set $t$ as $t_{\text{middle}}$ to solve Problem (\ref{TP33}), which indicates whether Problem (\ref{TP31}) is feasible or not and decides whether we should search in $[t_{\text{lower}}, t_{\text{middle}}]$ or $[t_{\text{middle}}, t_{\text{upper}}]$. Then we update the interval accordingly. The bisection process stops until some convergence criterion is met. The details are shown in Algorithm 2.

\renewcommand{\algorithmicensure}{\textbf{Iteration:}}

\begin{algorithm} 
\caption{Transmit Power Allocation Algorithm}
\label{alg:Resource allocation2} 

\begin{spacing}{1.2}\begin{algorithmic}[1]
\IF{$\mathcal{Q}_n > P_n^{max}$}
\RETURN Problem (\ref{TP2}) has no solution.
\ENDIF
\REQUIRE ~~\\
 $\bullet$  Set $ t_{\text{lower}} =0$, $ t_{\text{upper}} =F_n(P_n^{max}) = \frac{\gamma_n\varphi s_n^2 P_n^{max}}{B_n\log_2(1+\frac{P_n^{max} h_n}{B_n\sigma^2})} $.\\
 $\bullet$  Set iteration number $i =0$ and the maximum tolerance $\epsilon>0$.\\
\ENSURE ~~\\
\WHILE{$\mid t_{\text{upper}} - t_{\text{lower}} \mid\leq\epsilon$}
\STATE $i \leftarrow i+1$; 
\STATE Compute $t_{\text{middle}}=( t_{\text{lower}} + t_{\text{upper}} )/2$. 
\STATE Solve Problem (\ref{TP33}) with $t$ being $t_{\text{middle}}$ to obtain the optimal variable value $P_n^{(i)}$ and the optimal objective-function value $\phi_n^{(t_{\text{middle}})}(P_n^{(i)}) $.
\IF {$\phi_n^{(t_{\text{middle}})}(P_n^{(i)}) \leq 0$} \label{line8}
\STATE $ t_{\text{upper}} =t_{\text{middle}}$ and $case = 0$.\label{line9} //\textbf{Comment: } This case means $P_n^{(i)}$ is a feasible solution to Problem (\ref{TP31}) when $t$ is $t_{\text{middle}}$, so $t_{\text{middle}} \geq $ the optimal objective-function value of Problem (\ref{TP2}).  
\ELSE\label{line10} 
\STATE $ t_{\text{lower}} =t_{\text{middle}}$ and $case = 1$.\label{line11}  //\textbf{Comment: } This case means Problem (\ref{TP31}) is infeasible when $t$ is $t_{\text{middle}}$, so $t_{\text{middle}} < $ the optimal objective-function value of Problem (\ref{TP2}). 
\ENDIF
\ENDWHILE
\IF {$case = 0$}
\RETURN $P_n^{(i)}$ as the solution to Problem (\ref{TP2}). //\textbf{Comment: } Here $F_n(P_n^{(i)}) $ will be no greater than the current $t_{\text{upper}}$ because of Lines~\ref{line8} and~\ref{line9}, and hence $F_n(P_n^{(i)}) $ differs with the optimal objective-function value by at most $\epsilon$.
\ELSE
\RETURN $P_n^{\#}$ as the solution to Problem (\ref{TP2}), where $P_n^{\#}$ denotes the optimal variable value of Problem (\ref{TP33}) with $t$ being $t_{\text{upper}}$.
% $P_n^{(i)}$ if $F_n(P_n^{(i)}) \leq F_n(P_n^{(i-1)}) $ and \textbf{return} $P_n^{(i-1)}$ otherwise.
//\textbf{Comment: } Here the last iteration of the ``while'' loop makes Problem (\ref{TP31}) infeasible and we cannot just use $P_n^{(i)}$. For $F_n(P_n^{(i)}) $, we just know it is greater than the current $t_{\text{lower}}$ because of Lines~\ref{line10} and~\ref{line11}. Let $P_n^{*}$ be the (unknown) optimal solution of Problem (\ref{TP2}).
% Note that the current $t_{\text{upper}}$ is an upper bound the optimal objective-function value of Problem (\ref{TP2})).
We have $F_n(P_n^{\#}) \leq  t_{\text{upper}}$ since $\phi_n^{(t_{\text{upper}})}(P_n^{\#}) \leq \phi_n^{(t_{\text{upper}})}(P_n^{*}) \leq 0$, where $ \phi_n^{(t_{\text{upper}})}(P_n^{*}) \leq 0$ follows from  
$F_n(P_n^{*}) \leq t_{\text{upper}} $. Then $F_n(P_n^{\#})  $ differs with the optimal objective-function value by at most $\epsilon$.
% , the result of Problem (\ref{TP33}) with $t$ being $t_{\text{upper}}$ will make Problem (\ref{TP31}) 
\ENDIF
%\STATE Set $R_n(o)=B_n\log_2(1+\frac{P_nh_n}{B_n\sigma^2})$
% \STATE Compute $\upsilon(o)=\frac{\gamma_n\varphi s_n^2P_n}{B_n\log_2(1+\frac{P_nh_n}{B_n\sigma^2})} $
% \IF {$P_n(o)>P_n^{max}$ and $\upsilon(o)>\upsilon$}
% \STATE Solve (\ref{TP5}) to derive $P_{n}(o)=P_n^{max}$.
% \STATE Evaluate $R_n(o)=B_n\log_2(1+\frac{P_nh_n}{B_n\sigma^2})$.
% \STATE \textbf{break}.
% \ENDIF
\end{algorithmic}\end{spacing} 

\end{algorithm}

\subsection{CPU Frequency of AR Vehicle}
Given transmit power $\boldsymbol P$, computation model size $\boldsymbol s$, and computational resource allocation $\boldsymbol f^s$ on the Metaverse server, the CPU frequency of AR vehicle problem can be rewritten as
\begin{eqnarray}\label{P3}
&&\!\!\!\!\!\min\limits_{\boldsymbol f} \sum\limits_{n\in\mathcal{N}}\kappa_n f_n^2+\frac{\zeta C_n}{f_n} \nonumber\\
&&\!\!\!\!\!\mathrm{s.t.}~~\mathrm{(C2)}:~f_{min}\leq f_n\leq f_{max},~~\forall n, \nonumber \\
&&~~~\mathrm{(C6)}:~f_n\geq\frac{C_n}{\tau_{max}^{n}},~~\forall n,\nonumber\\
&&~~~\mathrm{(C7)'}: f_n>0,~~\forall n,
\end{eqnarray}
where $\tau_{max}^{n}\triangleq T-\frac{\varphi s_n^2}{R_n}-\frac{\varphi s_n^2 c_n}{f_n^s}$.

\renewcommand{\algorithmicensure}{\textbf{Iteration:}}

In (\ref{P3}), the objective function is convex and the constraints are linear, so we can derive that (\ref{P3}) is a convex optimization problem. Meanwhile, we also find that the constraints of (\ref{P3}) are for individual $f_n$. Then, its objective function can be dissolved into individual users. Consequently, the solution of (\ref{P3}) is obtained at either the minimum point of $\kappa_n f_n^2+\frac{\zeta C}{f_n}$ or at some boundary point. Then the optimal CPU frequency of an AR vehicle is given by 
\begin{eqnarray}\label{f}
f_n=\left\{\begin{array}{ll}
\mathfrak{F}_n, &\textrm{if $\mathfrak{D}_n\leq\mathfrak{F}_n\leq f_{max}$,}\\
\mathfrak{D}_n, &\textrm{if $\mathfrak{F}_n\leq \mathfrak{D}_n$,}\\
f_{max},&\textrm{if $\mathfrak{F}_n\geq f_{max}$,}\\
\end{array}\right.
\end{eqnarray}
where $\mathfrak{F}_n\triangleq\sqrt[3]{\frac{\zeta C_n}{2\kappa_n}}$ and $\mathfrak{D}_n\triangleq\max\{\frac{C_n}{\tau_{max}^{n}}, f_{min}\}$.
\subsection{Computational resource allocation on the server}\label{S32}
For a given transmit power $\boldsymbol P$, computation model size $\boldsymbol s$, and CPU frequency of AR vehicle $\boldsymbol f$, the computational resource allocation on the Metaverse server can be rewritten as
\begin{eqnarray}\label{P4}
&&\min\limits_{\boldsymbol f^s}~\sum\limits_{n\in \mathcal{N}}\beta_n\kappa_{ser}(f_n^s)^2\varphi s_n^2c_n\nonumber \\
&&\mathrm{s.t.}~\mathrm{(C3)}:~\sum\limits_{n\in\mathcal{N}}f_n^s\leq F, \\
&&~~~~~\mathrm{(C6)}:~\frac{\varphi s_n^2}{R_n}+\frac{\varphi s_n^2 c_n}{f_n^s}+T_n^{cv}\leq T,~~\forall n. \nonumber\\
&&~~~~~\mathrm{(C7)}:~f_n^s>0~~\forall n. \nonumber
\end{eqnarray}

Since both the objective function and the constraints of (\ref{P4}) are convex and linear, respectively, it is a convex optimization problem. Therefore, we can attain its solution by utilizing the Lagrangain dual decomposition technology. Thus, we first give its Lagrangain function as follows:
\begin{eqnarray}
&&\!\!\!\!\!\!\!\!\!\!\!\!L(\boldsymbol f^s, \nu, \boldsymbol \mu)=\sum\limits_{n\in\mathcal{N}}\beta_n\kappa_{ser}(f_n^s)^2\varphi s_n^2c_n+\nu(\sum\limits_{n\in\mathcal{N}}f_n^s-F) +\sum\limits_{n\in\mathcal{N}}\mu_n(\frac{\varphi s_n^2c_n}{T-T_n^{cv}-\frac{\varphi s_n^2}{R_n}}-f_{n}^s)
\end{eqnarray}
where $\boldsymbol \mu\triangleq(\mu_1, \mu_2, \ldots, \mu_N)\succeq0$ and $\nu\geq0$ are the Lagrangain parameters commensurate with $(\mathrm{C3})$ and $(\mathrm{C6})$ in (\ref{P4}), respectively.

Wherefore, the Lagrange dual function is given by
\begin{eqnarray}\label{L}
D(\nu, \boldsymbol \mu)=\min\limits_{\boldsymbol f^s\text{ satisfying }(\mathrm{C7})} L(\boldsymbol f^s, \nu, \boldsymbol \mu).
\end{eqnarray}
The dual function $D(\nu, \boldsymbol \mu)$ is concave since it is the minimum of linear functions.

Observing (\ref{L}), for a given Lagrange multipliers $\nu,\boldsymbol \mu$, we can minimize $L(\boldsymbol f^s, \nu, \boldsymbol \mu)$ to solve the following problem.
\begin{eqnarray}\label{L1}
&&\!\!\!\!\!\!\min\limits_{\boldsymbol f^s}~\sum\limits_{n\in\mathcal{N}}\beta_n\kappa_{ser}(f_n^s)^2\varphi s_n^2c_n+\nu(\sum\limits_{n\in\mathcal{N}}f_n^s-F)+\sum\limits_{n\in\mathcal{N}}\mu_n(\frac{\varphi s_n^2c_n}{T-T_n^{cv}-\frac{\varphi s_n^2}{R_n}}-f_{n}^s)\nonumber\\
&&\!\!\!\!\!\!\mathrm{s.t.}~(\mathrm{C7}):~~f_n^s>0, ~~\forall n,
\end{eqnarray}

In order to solve (\ref{L1}), we differentiate $L(\boldsymbol f^s, \nu, \boldsymbol \mu)$ with respect to $f_n^s$, and set $\frac{\partial L(\boldsymbol f^s, \nu, \boldsymbol \mu)}{\partial f_n^s}=0$. Then we have
\begin{eqnarray}\label{fs}
f_n^{s*}=\frac{\mu_n-\nu}{2\beta_n\kappa_{ser}\varphi s_n^2c_n}.
\end{eqnarray}

From (\ref{fs}), we can observe that to get value of $\boldsymbol f^s$ , the dual variables $(\nu, \boldsymbol \mu)$ need to be known in advance. Then the dual problem of (\ref{P4}) is expressed as
\begin{eqnarray}\label{D}
&&\min\limits_{\nu, \boldsymbol \mu} -D(\nu, \boldsymbol \mu)\nonumber\\
&&\mathrm{s.t.}~\nu\geq0,~\boldsymbol \mu\succeq0,
\end{eqnarray}
where we write ``minimizing $-D(\nu, \boldsymbol \mu)$'' instead of ``maximizing $D(\nu, \boldsymbol \mu)$'' since we will take advantage of the convexity of $-D(\nu, \boldsymbol \mu)$ to compute its subgradient.

In particular, the subgradient  method is adopted to tackle the dual problem. Therefore, we have the following theorem to compute the subgradient of $-D(\nu, \boldsymbol \mu)$. The proof is shown in Appendix B.
\begin{theorem} \label{themsubgradient}
The subgradient of $-D(\nu, \boldsymbol \mu)$ can be expressed as
\begin{eqnarray}
&&\!\!\!\!\!\!\!\!\!\! \frac{\partial \big(- D(\nu, \boldsymbol \mu) \big)}{\partial \nu} =F - \sum\limits_{n\in\mathcal{N}}f_n^{s*},~~\forall n,\\
&&\!\!\!\!\!\!\!\!\!\!\frac{\partial \big(-D(\nu, \boldsymbol \mu)\big)}{\partial \mu_n}=f_{n}^{s*}-\frac{\varphi s_n^2c_n}{T-T_n^{cv}-\frac{\varphi s_n^2}{R_n}},~~\forall n,
\end{eqnarray}
\mbox{where $f_{n}^{s*}$ is the optimal computational resource allocation on the Meteverse server, given $\nu$ and $\boldsymbol \mu$.} \end{theorem}

Then the proposed iterative algorithm to solve (\ref{P4}) is presented in Algorithm 3.
\begin{algorithm}[!t]
\caption{Iterative algorithm for computational resource allocation on the Metaverse server}
\label{alg:Resource allocation3}
\begin{spacing}{1.2}\begin{algorithmic}[1]
\REQUIRE ~~\\
 $\bullet$   Set $\nu$, $\boldsymbol \mu$, $t_{\text{upper}}$ and $\varsigma$.\\
 $\bullet$   Set $t=0$.
\ENSURE ~~\\
\WHILE {$t\leqslant t_{\text{upper}}$}
\STATE Compute CPU frequency $f_{n}^s(t)$ according to (\ref{fs}).
\STATE Update $\nu$ and $\boldsymbol \mu$, according to the following method, respectively,
\begin{eqnarray}
&&\nu(t+1)= \max\left\{0,~ \nu(t)-i(t) \cdot \frac{\partial \big(- D(\nu, \boldsymbol \mu) \big)}{\partial \nu} \right\} ,\\
&&\mu_n(t+1)=\max\left\{0,~\mu_n(t)-j(t) \cdot \frac{\partial \big(-D(\nu, \boldsymbol \mu)\big)}{\partial \mu_n} \right\},
\end{eqnarray}
where $i(t)$ and $j(t)$ are non-negative step size, and the expressions of the subgradients are given in Theorem~\ref{themsubgradient} above.
\IF {$|\nu(t+1)-\nu(t)|<\varsigma$ and $||\boldsymbol \mu(t+1)-\boldsymbol \mu(t)||_\textrm{2}<\varsigma$\\}
%\STATE \textbf{then}\\
\STATE $f_{n}^{s*}(t)=f_{n}^s(t)$.
\STATE \textbf{break}.\\
\ELSE
\STATE $t=t+1$.
\ENDIF
\ENDWHILE
\end{algorithmic}\end{spacing}
\end{algorithm}

Based on the above analysis, we develop the joint resource allocation policy for AR-empowered vehicular edge Metaverse system (AR), which is shown in Algorithm 4. First, the algorithm is initialized with the smallest computation model size, the lowest transmit power, the minimum CPU frequency of each AR vehicle, and evenly allocated computational resource on the server among AR vehicles. Afterwards, we update $\boldsymbol s$, $\boldsymbol P$, $\boldsymbol f$, and $\boldsymbol f^s$ sequentially until convergence (i.e., Lines 1--10 in Algorithm 4).

\begin{algorithm}[!t]
\caption{A Joint Algorithm for AR-Empowered Vehicular Edge Metaverse System}
\label{alg:Resource allocation4}
\begin{spacing}{1.2}\begin{algorithmic}[1]
\REQUIRE ~~\\
 $\bullet$  Set computation model size $\boldsymbol s(0)$, transmit power $\boldsymbol P(0)$, and CPU frequency of AR vehicle $\boldsymbol f(0)$. \\
 $\bullet$  Set $l=0$ and the maximum tolerance $\xi>0$.\\
\STATE Compute computational resource allocation 
 $\boldsymbol f^s(0)$ on the server by calling Algorithm 3 according to $\boldsymbol s(0)$, $\boldsymbol P(0)$, and $\boldsymbol f(0)$. \\
\STATE Compute $C(0)$ according to $\boldsymbol f(0)$, $\boldsymbol s(0)$, $\boldsymbol P(0)$, and $\boldsymbol f_n^s(0)$, which is expressed as
\begin{eqnarray}\label{C}
&&\!\!\!\!\!\!\!\!\!\!\!\!\!\!\!\!C(0)=\sum\limits_{n\in\mathcal{N}}\left(\rho_n\ln(1+\varphi s_n^2(0)-\beta_n\kappa_{ser}(f_n^s(0))^2\varphi s_n^2(0)c_n-\gamma_n\frac{P_n(0)\varphi s_n^2(0)}{B_n\log_2(1+\frac{P_n(0)h_n}{B_n\sigma^2})}\right)
\end{eqnarray}
\ENSURE ~~\\
\STATE Update $l=l+1$.\\
\STATE Obtain computation model size $\boldsymbol s(l)$ by calling Algorithm 1 given $\boldsymbol f(l-1)$, $\boldsymbol P(l-1)$, and $\boldsymbol f^s(l-1)$.
\STATE Allocate $\boldsymbol P(l)$ by calling Algorithm 2 given $\boldsymbol f(l-1)$, $\boldsymbol s(l)$, and $\boldsymbol f^s(l-1)$.
\STATE Compute CPU frequency $\boldsymbol f(l)$ of AR vehicle from (\ref{f}) given $\boldsymbol s(l)$, $\boldsymbol P(l)$, and $\boldsymbol f^s(l-1)$. 
\STATE Compute computational resource allocation $\boldsymbol f^s(l)$ on the server by calling Algorithm 3 given $\boldsymbol s(l)$, $\boldsymbol P(l)$, and $\boldsymbol f(l)$.
\STATE Obtain $C(l)$ from (\ref{C}).
\STATE \textbf{Until} $\mid (C(l)-C(l-1))/C(l)\mid\leq\xi$.
\end{algorithmic}\end{spacing}
\end{algorithm}
\section{Convergence and Complexity Analysis}\label{sec:41}
In this section, we analyze the convergence and complexity of the proposed algorithms. Based on the discussion in \cite{boyd2003subgradient}, the subgradient method with constant step size yields convergence to the optimal value. We can observe that the complexity of Algorithm 4 is mainly derived from Step 4 (i.e., Algorithm 1), Step 5 (i.e., Algorithm 2), and Step 7 (i.e., Algorithm 3), respectively. In Algorithm 1, the length of the search intervals is $M$, then its complexity is $O(\log_2(M))$. Since Algorithm 2 is essentially a bisection method, the number of iterations for the algorithm to converge is $O(\log_2(\frac{ t_{\text{upper}} - t_{\text{lower}} }{\epsilon}))$.
In Algorithm 3, the upper bound on the complexity of the subgradient  method is $O(\frac{1}{\varsigma^2})$ \cite{boyd2004convex}. The sum number of iterations of Algorithm 4 is assumed to be $\chi$. In the simulation, we can observe that the $\chi$ is typically no more than $7$. Consequently, the complexity of the algorithm is expressed as $O((\log_2(M)+\log_2(\frac{ t_{\text{upper}} - t_{\text{lower}} }{\epsilon})+\frac{1}{\varsigma^2})\chi)=O(7(\log_2(M)+\log_2(\frac{ t_{\text{upper}} - t_{\text{lower}} }{\epsilon})+\frac{1}{\varsigma^2}))$.

\section{Performance and Discussions} \label{sec:4}
In this section, a plethora of simulation results is particularized to estimate the performance of the proposed algorithms. Primarily, we give the settings of the relevant parameter values. Subsequently, the convergence of the proposed algorithms is verified by the simulation. Finally, we give three baseline schemes, which will be established next, to demonstrate the performance of the proposed scheme.

\subsection{Simulation Parameters}
\begin{figure*}[tp]
\begin{minipage}[t]{0.5\linewidth}
\centering
\includegraphics[height=2.8in,width=8.0cm]{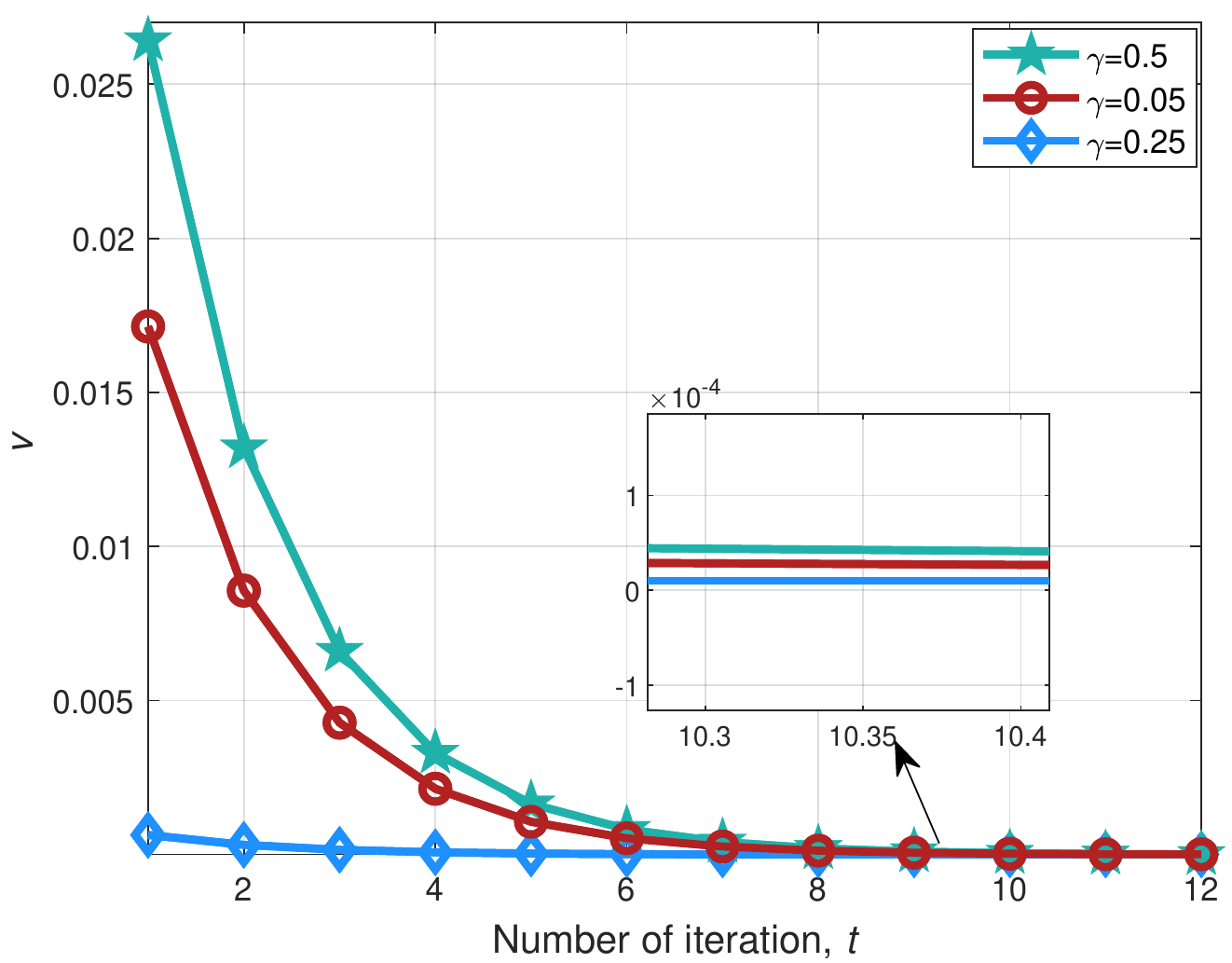}
\caption{\small Convergence of Algorithm 2.}
\label{fig:2}
\end{minipage}%
\begin{minipage}[t]{0.5\linewidth}
\centering
\includegraphics[height=2.8in,width=8.0cm]{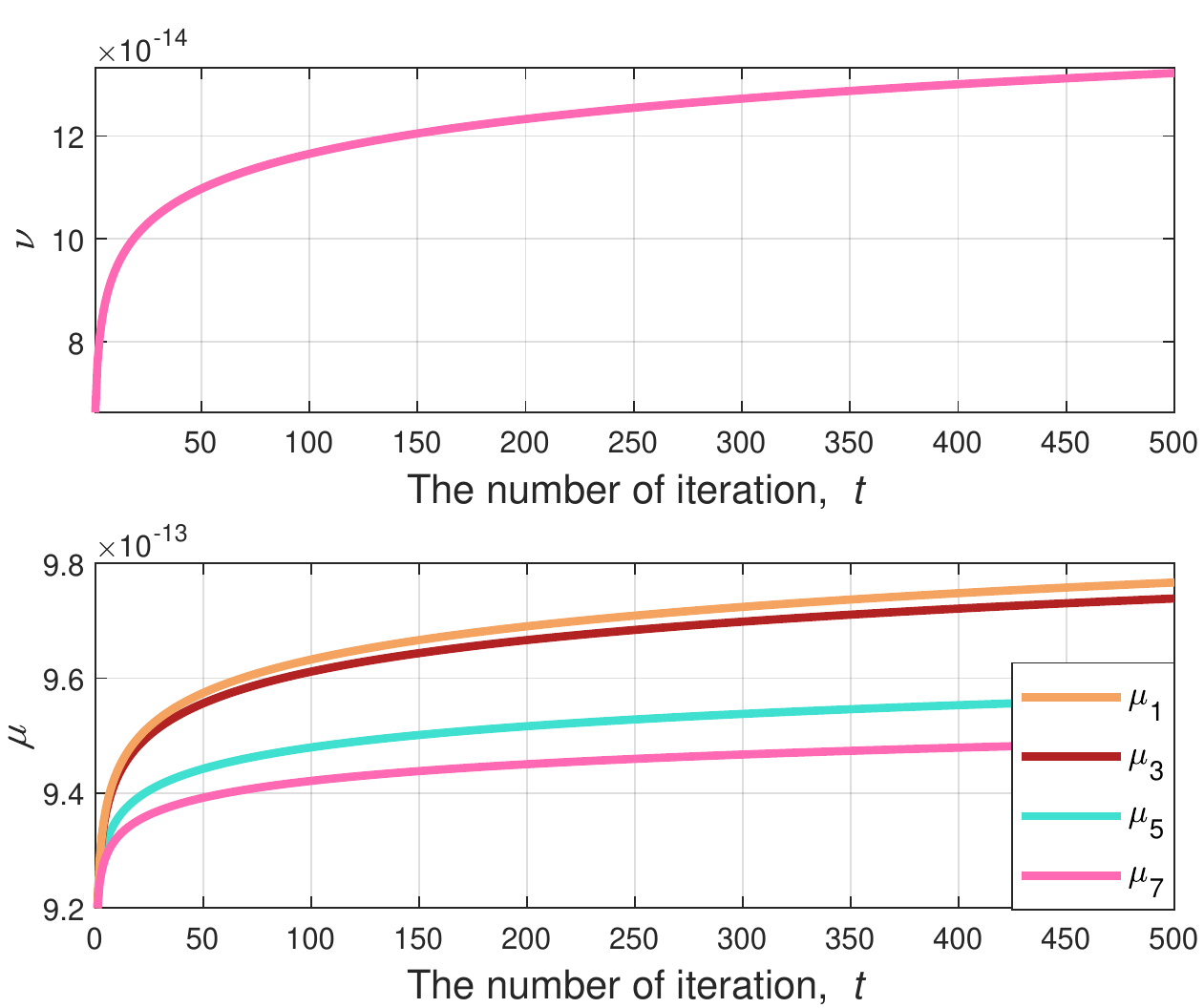}
\caption{\small Convergence of Algorithm 3.}
\label{fig:3}
\end{minipage}
\end{figure*}
We consider an AR-empowered vehicular edge Metaverse system consisting of a VEC server and $30$ AR vehicles over a $5\times5$ km area, where each vehicle travels at a constant speed. Similar to \cite{9745059}, the channel between BS and AR vehicles is modeled as $h=\theta d^{-3} \bar{h}$, where $\theta$ is a constant ($\theta=32$ dB), $d$ denotes the distance between the BS and AR vehicles, and $\bar{h}$ follows Gaussian distribution with zeros mean and covariance $\textbf{\textit{I}}$. The minimum computation model size is $10000$ pixels ($100\times100$). The default value of $\rho_n$ is $0.04$ \cite{li2018wirelessly}.
Other simulation parameter settings are shown in Table II.

\begin{table}
\centering \small
\caption{Summary of Simulation Parameters}
\begin{tabular}{ll}
  \toprule[2pt]
  \\[-2mm]
 Parameters & Values \\
  \hline
  \vspace{0.8mm}
  System bandwidth, $B$ & $10$ MHz\\
  \hline
  \vspace{0.8mm}
  Total computing capacity of server, $F$ & $2.5$ GHz  \cite{feng2019cooperative}\\
  \hline
  \vspace{0.8mm}
 Computational capability of CPU, $f_{max}/f_{min}$ & $2/0.5$ GHz \\
 \hline
  \vspace{0.8mm}
  Block length, $T$ &  $1$s \\
  \hline
  \vspace{0.8mm}
 Processing density, $c_n$ & $737.5$ cycle/bit \cite{mao2017stochastic} \\
  \hline
  \vspace{0.8mm}
  Information carried $\varphi$ & $24$ \cite{ahn2019novel}\\
   \hline
  \vspace{0.8mm}
  Transmit power, $P_{n}^{max}$ & $2$ W\\
  \hline
  \vspace{0.8mm}
  Minimum analytics accuracy, $\delta_{n}$ & $0.85$ \\
  \hline
  \vspace{0.8mm}
   Effective switched capacitance, $\kappa_n, \kappa_{ser}$ & $10^{-27}$ \\
  \bottomrule[2pt]
\end{tabular}
\end{table}

%\begin{figure}[h!]
%\vspace*{-2mm}
%\begin{center}
%\includegraphics[width=0.28\textwidth]{convergence.eps}
%\end{center}
%\vspace*{-3mm}
%\caption{Convergence of Algorithm 1.}
%\label{Fig1}
%\vspace*{0mm}
%\end{figure}

%\begin{figure}[h!]
%\vspace*{-2mm}
%\begin{center}
%\includegraphics[width=0.28\textwidth]{operator_reward.eps}
%\end{center}
%\vspace*{-3mm}
%\caption{Operator's reward vs. maximum amount of transferred power.}
%\label{Fig1}
%\vspace*{0mm}
%\end{figure}

%\begin{figure}[h!]
%\vspace*{-2mm}
%\begin{center}
%\includegraphics[width=0.28\textwidth]{operator_reward1.eps}
%\end{center}
%\vspace*{-3mm}
%\caption{Operator's reward vs. input data size.}
%\label{Fig1}
%\vspace*{0mm}
%\end{figure} \underline{}

To demonstrate the performance of the proposed scheme, we set up three baselines, namely \underline{f}ixed \underline{c}omputation \underline{m}odel \underline{s}ize scheme (FCMS), \underline{f}ixed \underline{s}erver \underline{c}om\underline{p}utational resource allocation  (FSCP), and \underline{f}ixed \underline{C}PU frequency of \underline{AR} vehicles scheme (FARC). Details are as follows:
\begin{itemize}
\item FCMS: The scheme does not optimize the computation model size of AR vehicles $\boldsymbol s$.
\item FSCP: The scheme performs joint optimization over the CPU frequency of AR vehicle $\boldsymbol f$, the transmit power for communication model $\boldsymbol P$, and the computation model size of AR vehicle $\boldsymbol s$ except for computational resource allocation $\boldsymbol f^s$ on the Metaverse server.
\item FARC: The scheme jointly optimizes the computational resource allocation on the Metaverse server $\boldsymbol f^s$, the transmit power for communication model $\boldsymbol P$, and the computation model size of AR vehicle $\boldsymbol s$.
\end{itemize}

\subsection{Convergence Performance}
\begin{figure*}[tp]
\begin{minipage}[t]{0.5\linewidth}
\centering
\includegraphics[height=2.8in,width=8.0cm]{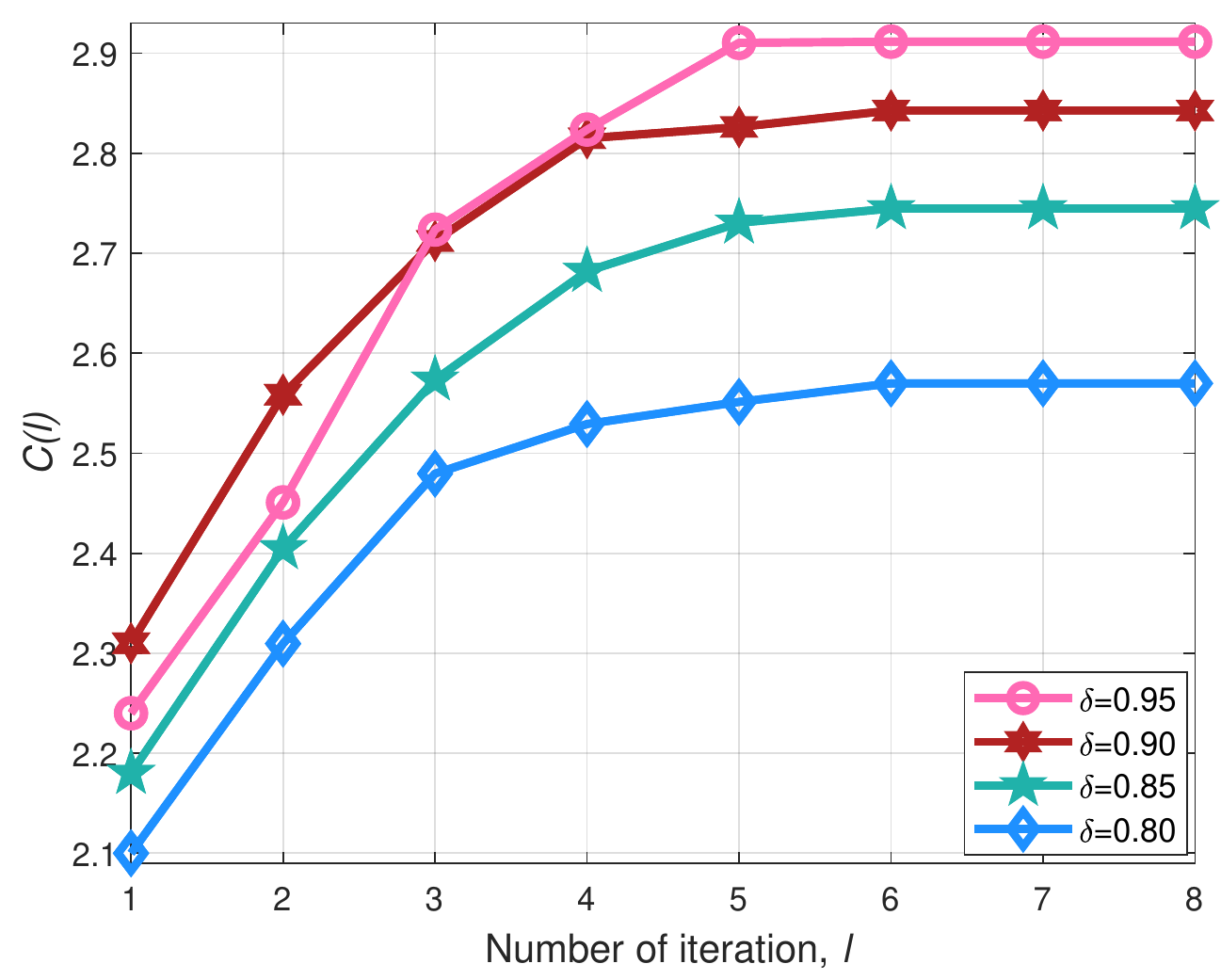}
\caption{\small Convergence of Algorithm 4.}
\label{fig:4}
\end{minipage}%
\begin{minipage}[t]{0.5\linewidth}
\centering
\includegraphics[height=2.8in,width=8.0cm]{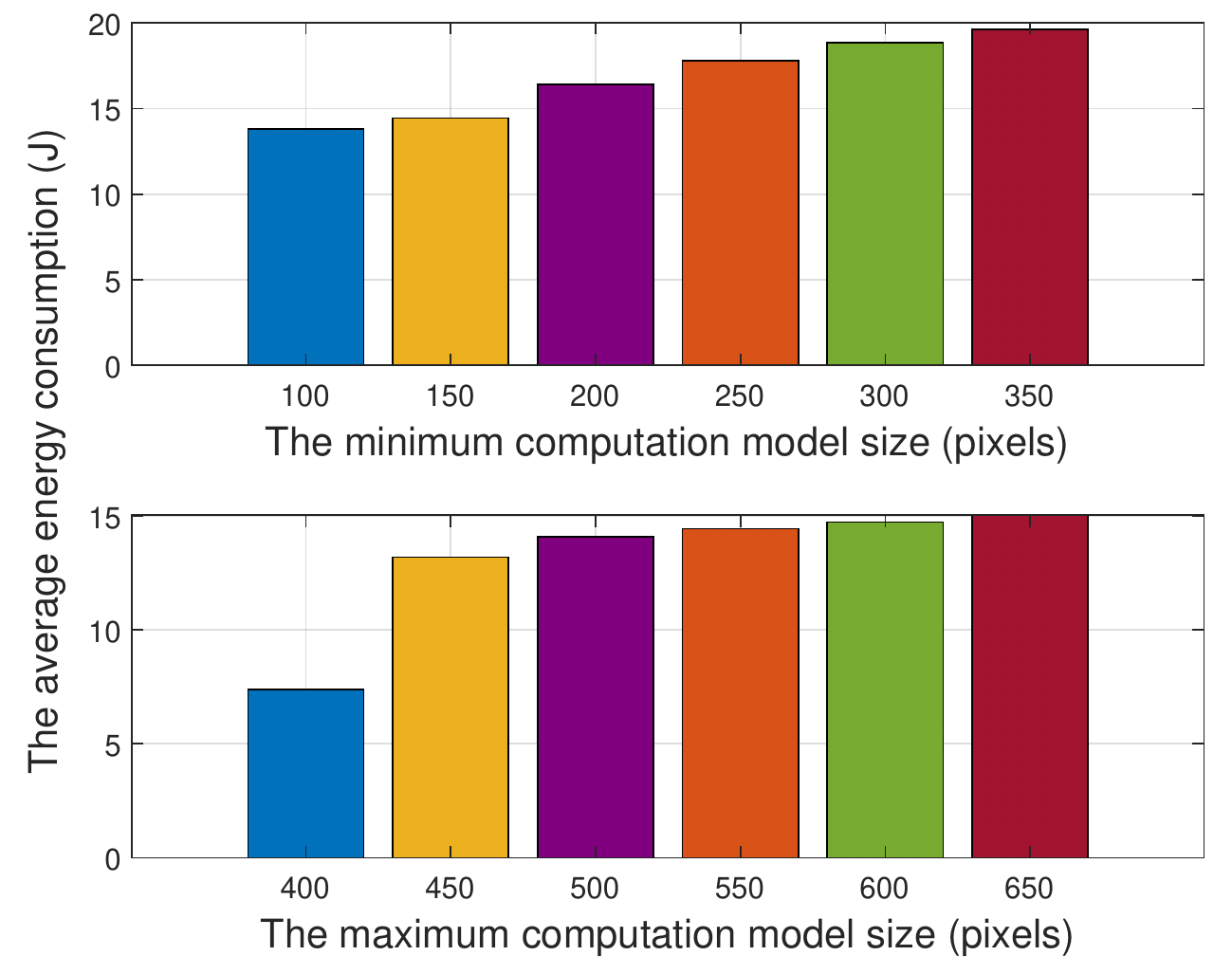}
\caption{\small Average energy consumption v.s. computation model size.}
\label{fig:5}
\end{minipage}
\end{figure*}
\begin{figure*}[tp]
\begin{minipage}[t]{0.5\linewidth}
\centering
\includegraphics[height=2.8in,width=8.0cm]{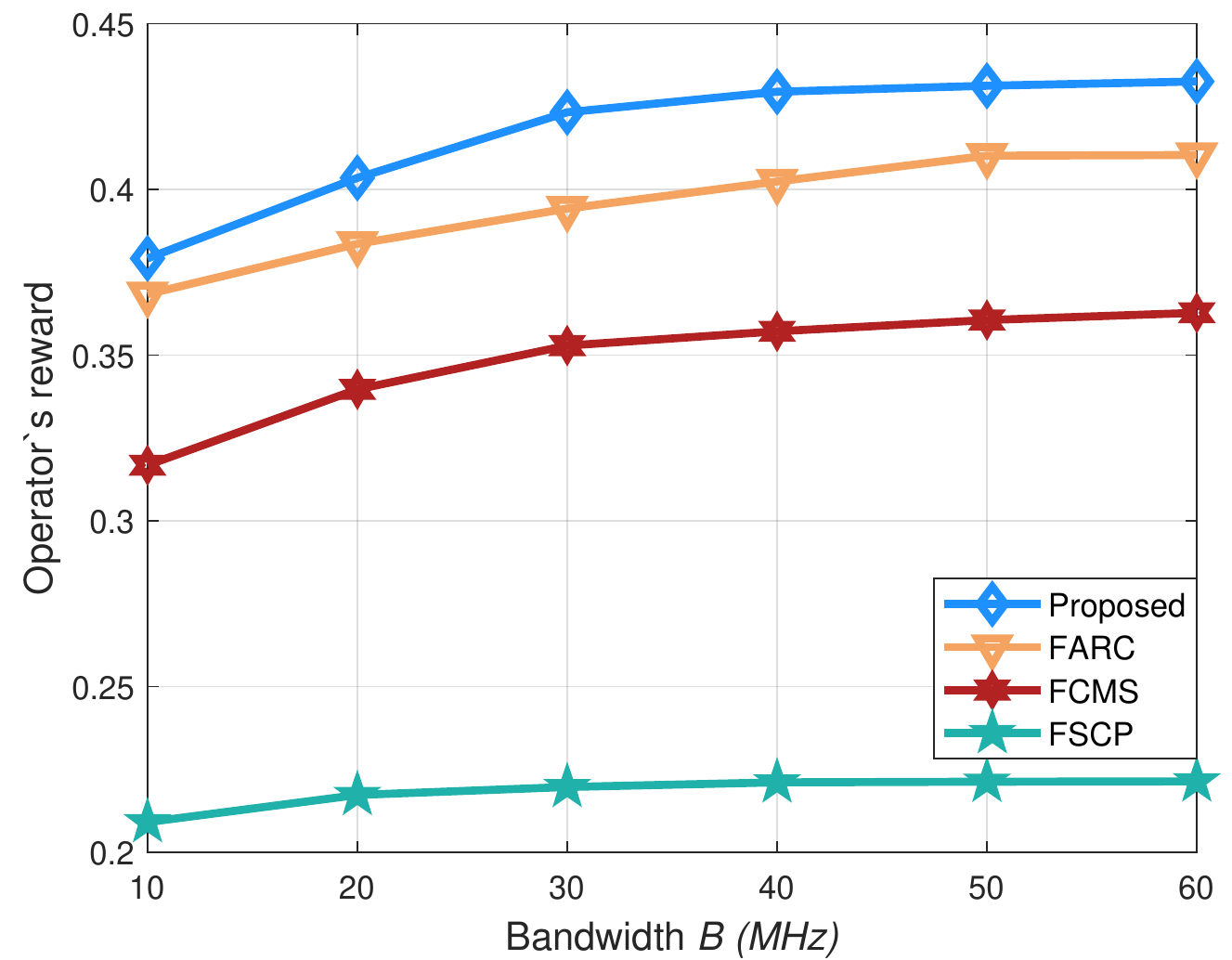}
\caption{\small System utility versus bandwidth $B$.}
\label{fig:6}
\end{minipage}%
\begin{minipage}[t]{0.5\linewidth}
\centering
\includegraphics[height=2.8in,width=8.0cm]{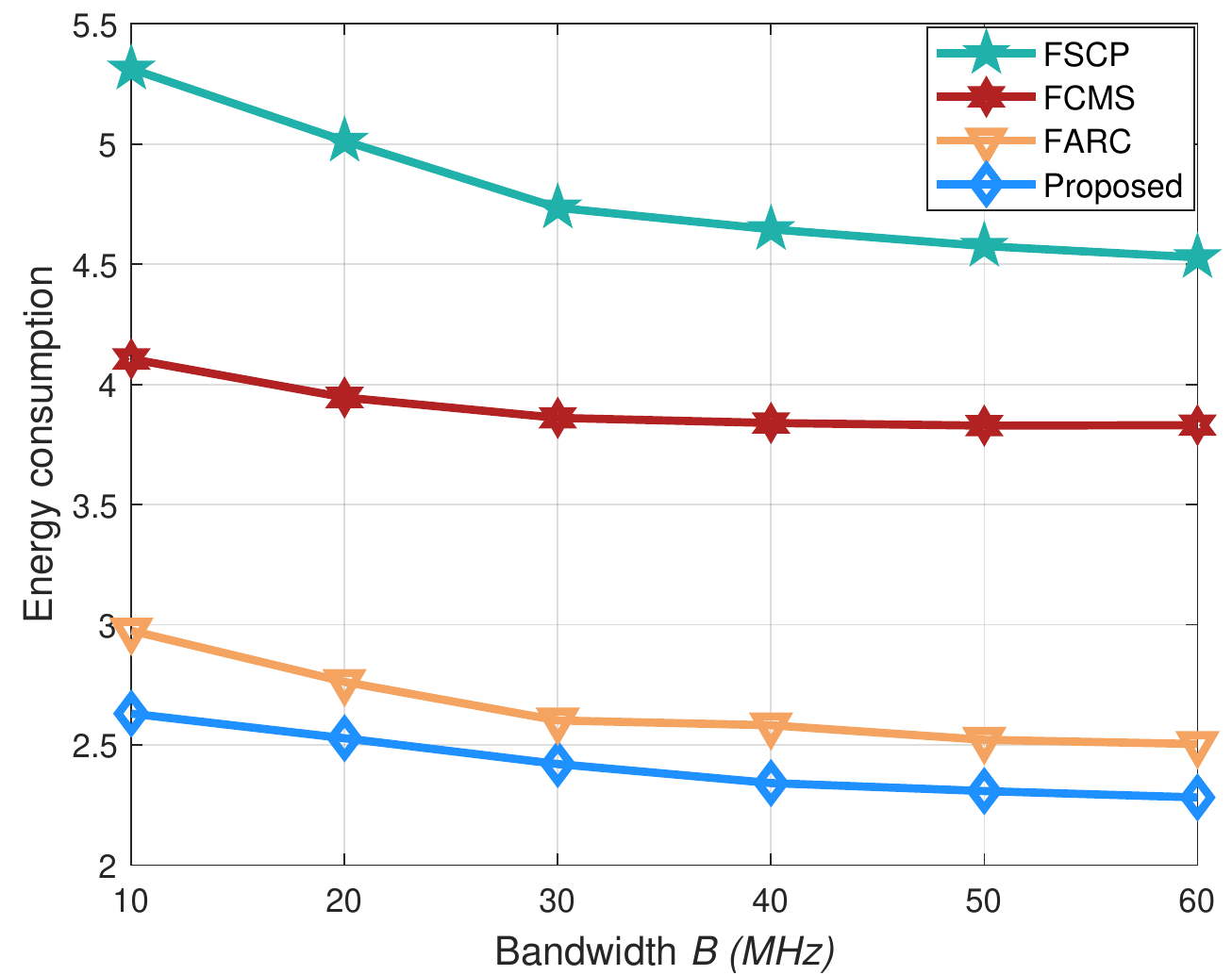}
\caption{\small Energy consumption versus bandwidth $B$.}
\label{fig:7}
\end{minipage}
\end{figure*}
In Fig. \ref{fig:2}, we manifest the convergence of Algorithm 2 and compare its convergence performance under different parameters $\gamma$. From the figure, we can observe that it converges at about the eighth step. Meanwhile, the smaller the value of $\gamma$, the faster the convergence rate. The reason is that the range of the bisection method is small, and the speed of obtaining the solution is fast. Fig. \ref{fig:3} unveils the convergence of Algorithm 3 by mapping the dual variables $\nu$ and $\boldsymbol \mu$, respectively, versus the number of iterations $t$. As shown in the figure, we can derive that the convergence rate of the algorithm is relatively fast. In Fig.3, the convergence of the dual variable $\boldsymbol \mu$ is obtained by randomly selecting four AR vehicles. In Fig. \ref{fig:4}, we show the convergence of Algorithm 4 to estimate the complexity of the entire scheme. From the figure, we can see that it has a fast convergence rate. Therefore, we can conclude that the proposed algorithm can solve the problem (\ref{P1}) efficiently.
\subsection{Performance of the Proposed Algorithm}
In Figs. \ref{fig:5}, we reveal the effect of minimum/maximum computation model size on the energy consumption. From the figure, we can find that the energy consumption becomes incremental with the increase in minimum/maximum computation model size $s_{min}/s_{max}$, which is in line with our intuition. A large computing model size will increase the transmission energy consumption of AR vehicles and increase the energy consumption of the server processing.

\begin{figure*}[tp]
\begin{minipage}[t]{0.5\linewidth}
\centering
\includegraphics[height=2.8in,width=8.0cm]{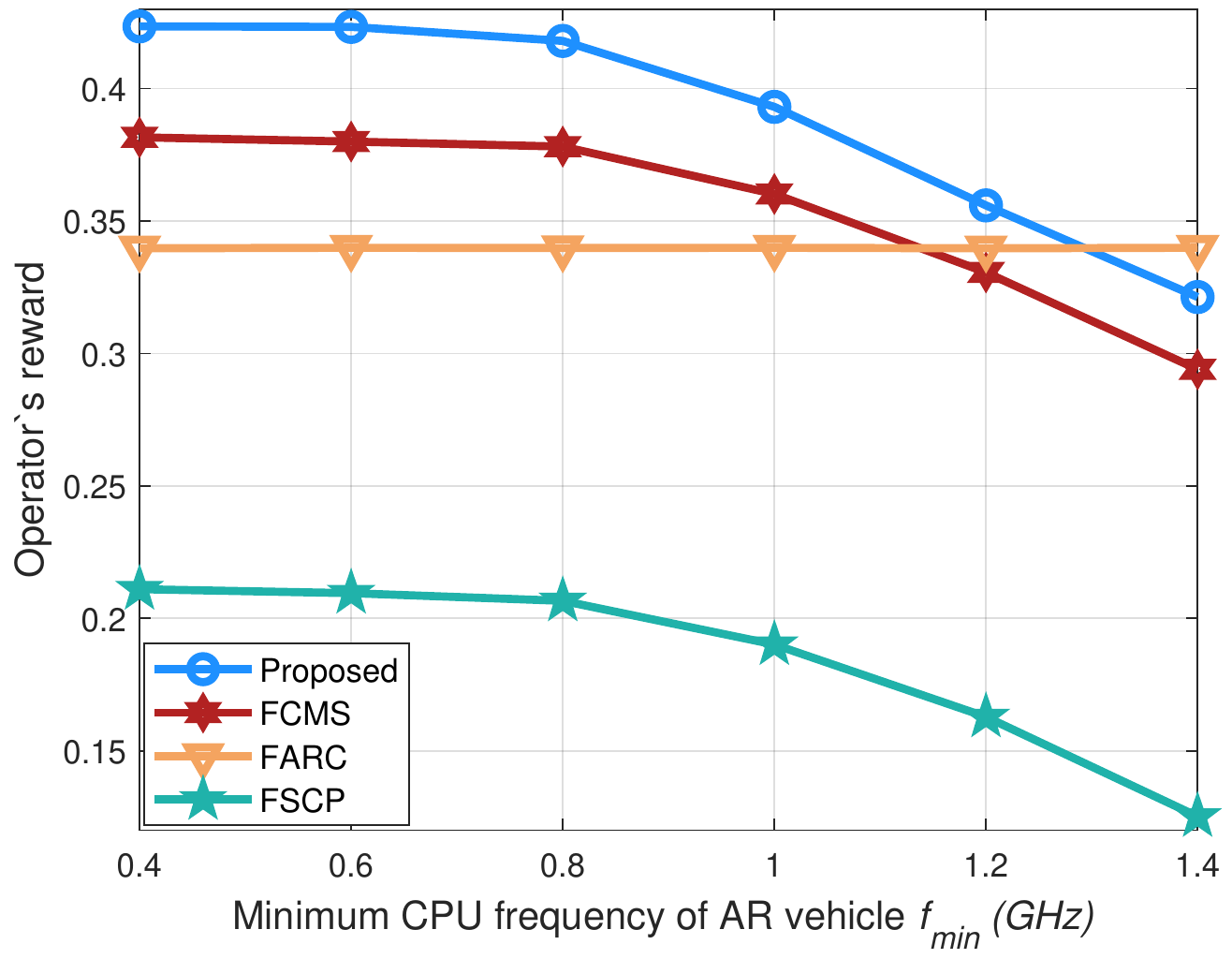}
\caption{\small System utility versus minimum CPU \\ frequency of AR vehicle $f_{min}$.}
\label{fig:8}
\end{minipage}%
\begin{minipage}[t]{0.5\linewidth}
\centering
\includegraphics[height=2.8in,width=8.0cm]{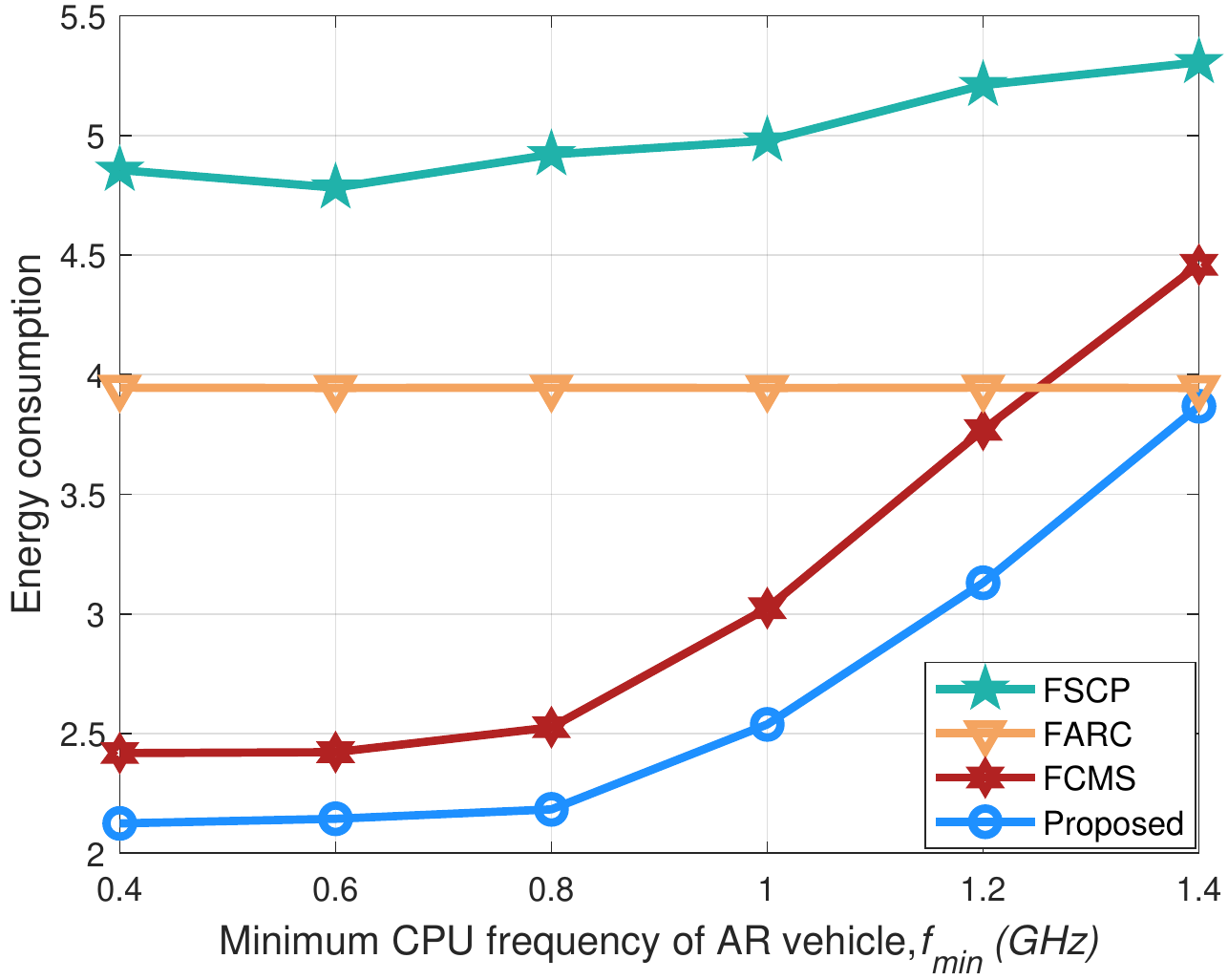}
\caption{\small Per frame energy consumption versus minimum CPU frequency of AR vehicle $f_{min}$.}
\label{fig:9}
\end{minipage}
\end{figure*}
\begin{figure*}[tp]
\begin{minipage}[t]{0.5\linewidth}
\centering
\includegraphics[height=2.8in,width=8.0cm]{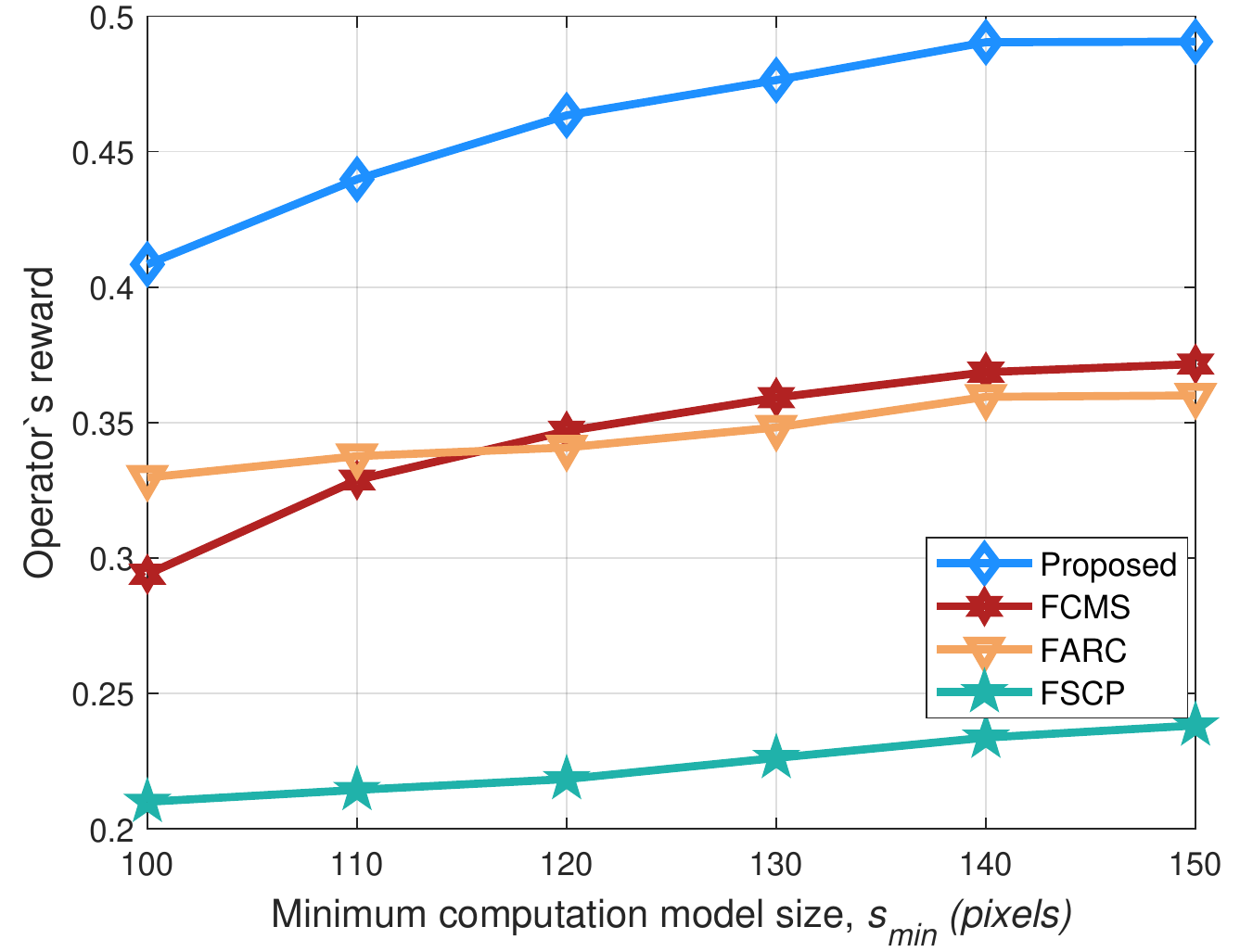}
\caption{\small System utility versus minimum computation model size $s_{min}$.}
\label{fig:10}
\end{minipage}%
\begin{minipage}[t]{0.5\linewidth}
\centering
\includegraphics[height=2.8in,width=8.0cm]{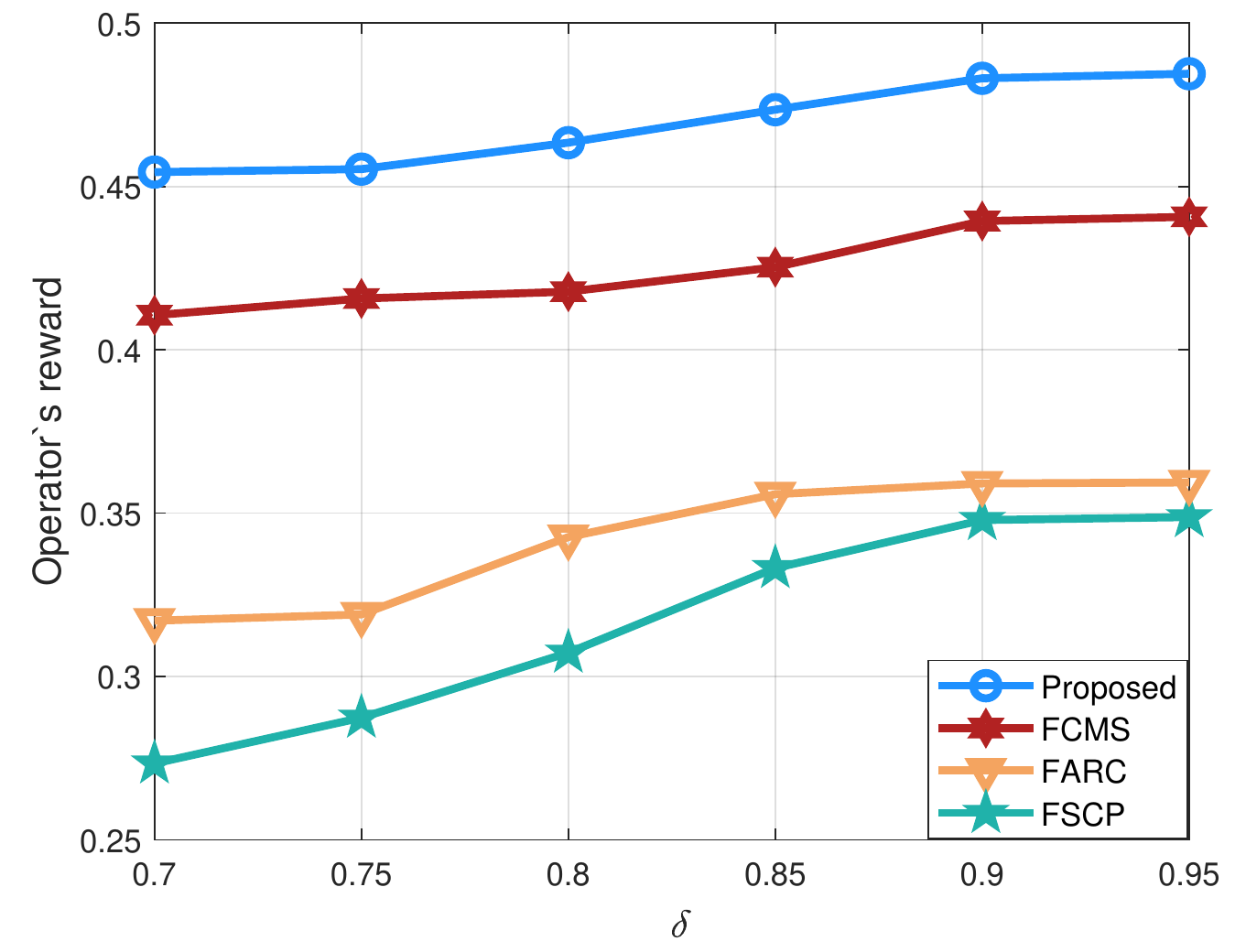}
\caption{\small System utility versus $\delta$.}
\label{fig:11}
\end{minipage}
\end{figure*}
In Figs. \ref{fig:6} and \ref{fig:7}, we manifest the system utility $C$ and the sum of  energy consumption ($\boldsymbol E^{ser}$, $\boldsymbol E^{cv}$, and $\boldsymbol E^{com}$) versus bandwidth $B$, respectively. As shown in Fig. 6, we can find out that the system utility is increasing leisurely with the increase in bandwidth $B$. It is easy to understand this phenomenon. When the computation model size is fixed, the increase in bandwidth leads to an increase in the transmission rate, which affects its energy consumption. In addition, we can observe that the proposed scheme outperforms always the best, followed by FARC, FCMS, and FSCP, successively.
\begin{figure*}[tp]
\begin{minipage}[t]{0.5\linewidth}
\centering
\includegraphics[height=2.8in,width=8.4cm]{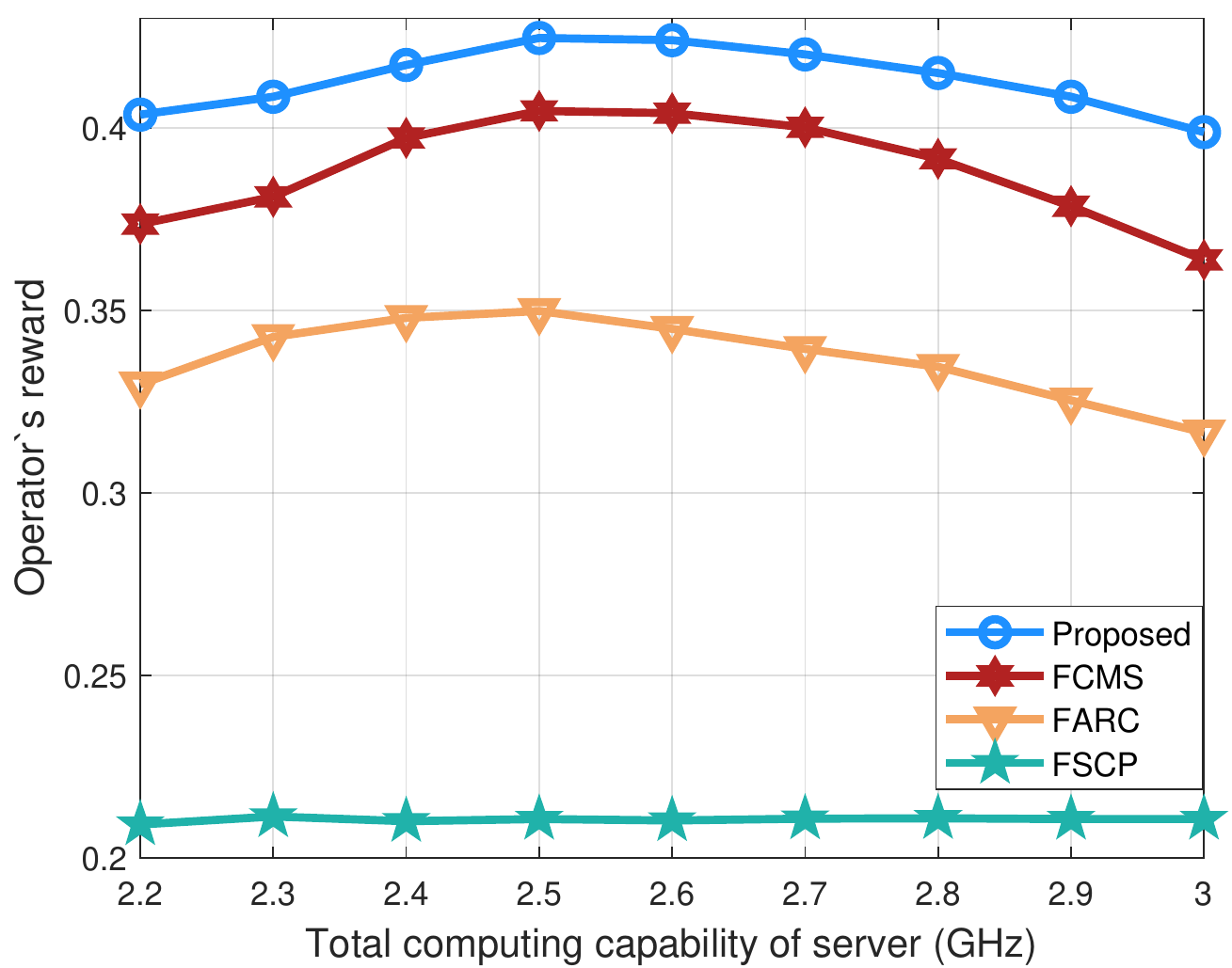}
\caption{\small System utility versus total computing \\ capability of server.}
\label{fig:12}
\end{minipage}%
\begin{minipage}[t]{0.5\linewidth}
\centering
\includegraphics[height=2.8in,width=8.4cm]{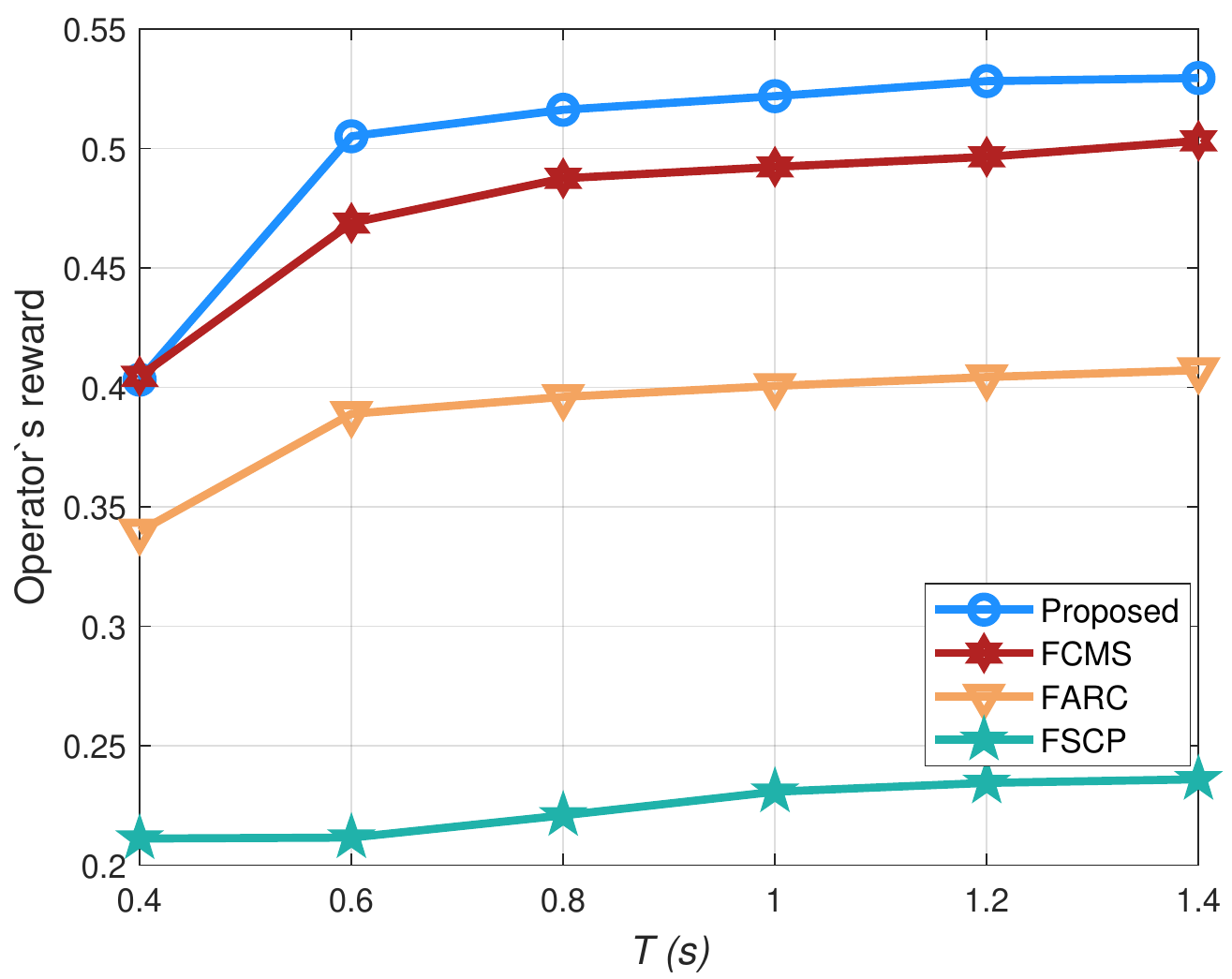}
\caption{\small System utility versus time block length $T$.}
\label{fig:13}
\end{minipage}
\end{figure*}
In Fig.~\ref{fig:7}, we show the performance of the proposed scheme under different bandwidths. As we can see that the energy consumption becomes smaller with the increase in bandwidth because the transmission energy is decreasing. From the figure, the energy consumption remains balanced when $B$ increases to a certain level (e.g., $B=40$ MHz). The reason is that the reduction in transmission energy has little effect on the overall energy consumption of the system for a given computation model size. Combined with Fig. \ref{fig:6}, we can infer that the proposed scheme reduces the energy consumption per frame by $18\%$ and increases the revenue by $22\%$ compared with the scheme FARC.

Fig. \ref{fig:8} show the impact of different minimum CPU frequency of AR vehicle $f_{min}$ on the system utility. From (1), we can know that the energy consumption is proportional to the CPU frequency, so the energy consumption becomes high while increasing the minimum CPU frequency of AR vehicle $f_{min}$ (see Fig. \ref{fig:9}), which will lead to less system utility. In particular, when the minimum CPU frequency of AR vehicles increases to a certain value (i.e., $f_{min}=0.8$ GHz), the system utility drops rapidly. The tendency shows that the value of the minimum CPU frequency has a significant impact on the system, and an appropriate minimum CPU frequency selection is critical to the system scheme. Meanwhile, we can also observe that the performance of FARC is almost unchanged, which means that the minimum CPU frequency does not determine the CPU frequency of the AR vehicles allocation strategy of FARC. Hence, its per energy consumption also does not convert, as shown in Fig. \ref{fig:9}.

In Fig. \ref{fig:10}, we exhibit the system utility versus the minimum computation model size $s_{min}$. From the operator's point of view, the more computation model size it gets, the more benefits it realizes. Therefore, the system utility becomes large while increasing the minimum computation model size $s_{min}$. From the figure, we can see that the proposed scheme outperforms other schemes, which indicates that the joint optimization of the CPU frequency of AR vehicle $\boldsymbol f$, the transmit power for communication model $\boldsymbol P$, the computation model size of AR vehicle $\boldsymbol s$, and the computational resource allocation $\boldsymbol f^s$ on the Metaverse server can enhance the performance of the system for a given minimum computation model size $s_{min}$.

To illustrate the impact of accuracy on the system performance, we show the system utility versus the minimum analytics accuracy requirement $\delta$ as shown in Fig. \ref{fig:11}. From the figure, we can observe that the system utility of all schemes is increasing with the increase in $\delta$. We can explain this reason in terms of constraint $(C5)$. Since the function $\ln(1-x)$ is strictly decreasing at $x\in[0,1]$, the function decrease as the value of $x$ increases. Therefore, the right-hand side of $(C5)$ increases as $\delta$ increases, which results in a more significant allocation of the computation model size. In addition, we can observe that the proposed scheme consistently outperforms the best, followed by FCMS, FARC, and FSCP, successively.

In Fig. \ref{fig:12}, we reveal the system utility versus the total computing capability of server $F$. We can see that when $F$ is less than a certain value (i.e., $F=2.6$ GHz), the system utility becomes larger with the increase in the total computing capability of server $F$. However, when $F$ exceeds $2.6$ GHz, the performance of all schemes degrades except for FSCP. The system performance is degraded because the increment of $F$ increases the energy consumption of the server. Since the allocation of computational resource on the server in FSCP is fixed, the increase of $F$ does not affect its performance, but its performance is the worst among all the schemes. Simultaneously, from the figure, we can also observe that the performance of the proposed scheme is always the best no matter how $F$ varies.

Fig. \ref{fig:13} manifests the system utility versus time block $T$. The minimum analytics accuracy requirement of AR vehicles is fixed as $0.85$ with the total computing capability of the server $F=2.5$ GHz. From the figure, it is observed that the system utility increases with the increase in time block length $T$. Meanwhile, the proposed scheme outperforms other schemes that randomly assign the value of $T$. Compared with other schemes, FSCP has the worst performance. Therefore, this indicates that the joint optimization of the CPU frequency of AR vehicle $\boldsymbol f$, the transmit power for communication model $\boldsymbol P$, the computation model size of AR vehicle $\boldsymbol s$, and the computational resource allocation $\boldsymbol f^s$ on the Metaverse server can enhance the system utility.
\section{Conclusion} \label{sec:5}
In this paper, we have proposed a resource allocation framework for augmented reality empowered vehicular edge Metaverse. The optimization framework can achieve the joint control of the CPU frequency of AR vehicles, the transmit power for the communication model, the computation model size of AR vehicles, and the computational resource allocation on the Metaverse server for maximizing the system utility. It provided insights into the control strategies of the Metaverse operator and the AR vehicles. Specifically, for the Metaverse operator, we manifested that it should designate AR vehicles to participate in data sharing operations based on their configuration and various parameters of the AR vehicles. For AR vehicles, each engagement should minimize its energy consumption as much as possible. In addition, the simulation results have exhibited that the proposed algorithm has good convergence characteristics. 

\begin{spacing}{1.18}
% \bibliographystyle{IEEEtran}
% \bibliography{Reference}

\begin{thebibliography}{10}
\providecommand{\url}[1]{#1}
\csname url@samestyle\endcsname
\providecommand{\newblock}{\relax}
\providecommand{\bibinfo}[2]{#2}
\providecommand{\BIBentrySTDinterwordspacing}{\spaceskip=0pt\relax}
\providecommand{\BIBentryALTinterwordstretchfactor}{4}
\providecommand{\BIBentryALTinterwordspacing}{\spaceskip=\fontdimen2\font plus
\BIBentryALTinterwordstretchfactor\fontdimen3\font minus
  \fontdimen4\font\relax}
\providecommand{\BIBforeignlanguage}[2]{{%
\expandafter\ifx\csname l@#1\endcsname\relax
\typeout{** WARNING: IEEEtran.bst: No hyphenation pattern has been}%
\typeout{** loaded for the language `#1'. Using the pattern for}%
\typeout{** the default language instead.}%
\else
\language=\csname l@#1\endcsname
\fi
#2}}
\providecommand{\BIBdecl}{\relax}
\BIBdecl

\bibitem{ning2021survey}
H.~Ning, H.~Wang, Y.~Lin, W.~Wang, S.~Dhelim, F.~Farha, J.~Ding, and
  M.~Daneshmand, ``A survey on metaverse: the state-of-the-art, technologies,
  applications, and challenges,'' \emph{arXiv preprint arXiv:2111.09673}, 2021.

\bibitem{cai2022compute}
Y.~Cai, J.~Llorca, A.~M. Tulino, and A.~F. Molisch, ``Compute-and
  data-intensive networks: The key to the metaverse,'' \emph{arXiv preprint
  arXiv:2204.02001}, 2022.

\bibitem{dowling2022non}
M.~Dowling, ``Is non-fungible token pricing driven by cryptocurrencies?''
  \emph{Finance Research Letters}, vol.~44, p. 102097, 2022.

\bibitem{zhou2019enhanced}
P.~Zhou, W.~Zhang, T.~Braud, P.~Hui, and J.~Kangasharju, ``Enhanced augmented
  reality applications in vehicle-to-edge networks,'' in \emph{2019 22nd
  Conference on Innovation in Clouds, Internet and Networks and Workshops
  (ICIN)}.\hskip 1em plus 0.5em minus 0.4em\relax IEEE, 2019, pp. 167--174.

\bibitem{braud2022scaling}
T.~Braud, C.~B. Fern{\'a}ndez, and P.~Hui, ``Scaling-up ar: University campus
  as a physical-digital metaverse,'' in \emph{2022 IEEE Conference on Virtual
  Reality and 3D User Interfaces Abstracts and Workshops (VRW)}.\hskip 1em plus
  0.5em minus 0.4em\relax IEEE, 2022, pp. 169--175.

\bibitem{liu2019edge}
L.~Liu, H.~Li, and M.~Gruteser, ``Edge assisted real-time object detection for
  mobile augmented reality,'' in \emph{The 25th Annual International Conference
  on Mobile Computing and Networking}, 2019, pp. 1--16.

\bibitem{9686591}
L.~Liu, M.~Zhao, M.~Yu, M.~A. Jan, D.~Lan, and A.~Taherkordi, ``Mobility-aware
  multi-hop task offloading for autonomous driving in vehicular edge computing
  and networks,'' \emph{IEEE Transactions on Intelligent Transportation
  Systems}, pp. 1--14, 2022.

\bibitem{sonmez2020machine}
C.~Sonmez, C.~Tunca, A.~Ozgovde, and C.~Ersoy, ``Machine learning-based
  workload orchestrator for vehicular edge computing,'' \emph{IEEE Transactions
  on Intelligent Transportation Systems}, vol.~22, no.~4, pp. 2239--2251, 2020.

\bibitem{dhelim2022edge}
S.~Dhelim, T.~Kechadi, L.~Chen, N.~Aung, H.~Ning, and L.~Atzori, ``Edge-enabled
  metaverse: The convergence of metaverse and mobile edge computing,''
  \emph{arXiv preprint arXiv:2205.02764}, 2022.

\bibitem{cai2022joint}
Y.~Cai, J.~Llorca, A.~M. Tulino, and A.~F. Molisch, ``Joint
  compute-caching-communication control for online data-intensive service
  delivery,'' \emph{arXiv preprint arXiv:2205.01944}, 2022.

\bibitem{duan2021metaverse}
H.~Duan, J.~Li, S.~Fan, Z.~Lin, X.~Wu, and W.~Cai, ``Metaverse for social good:
  A university campus prototype,'' in \emph{Proceedings of the 29th ACM
  International Conference on Multimedia}, 2021, pp. 153--161.

\bibitem{huynh2022artificial}
T.~Huynh-The, Q.-V. Pham, X.-Q. Pham, T.~T. Nguyen, Z.~Han, and D.-S. Kim,
  ``Artificial intelligence for the metaverse: A survey,'' \emph{arXiv preprint
  arXiv:2202.10336}, 2022.

\bibitem{siriwardhana2021survey}
Y.~Siriwardhana, P.~Porambage, M.~Liyanage, and M.~Ylianttila, ``A survey on
  mobile augmented reality with 5g mobile edge computing: architectures,
  applications, and technical aspects,'' \emph{IEEE Communications Surveys \&
  Tutorials}, vol.~23, no.~2, pp. 1160--1192, 2021.

\bibitem{liu2018dare}
Q.~Liu and T.~Han, ``Dare: Dynamic adaptive mobile augmented reality with edge
  computing,'' in \emph{2018 IEEE 26th International Conference on Network
  Protocols (ICNP)}.\hskip 1em plus 0.5em minus 0.4em\relax IEEE, 2018, pp.
  1--11.

\bibitem{redmon2017yolo9000}
J.~Redmon and A.~Farhadi, ``Yolo9000: better, faster, stronger,'' in
  \emph{Proceedings of the IEEE conference on computer vision and pattern
  recognition}, 2017, pp. 7263--7271.

\bibitem{wang2020user}
H.~Wang and J.~Xie, ``User preference based energy-aware mobile ar system with
  edge computing,'' in \emph{IEEE INFOCOM 2020-IEEE Conference on Computer
  Communications}.\hskip 1em plus 0.5em minus 0.4em\relax IEEE, 2020, pp.
  1379--1388.

\bibitem{du2021optimal}
H.~Du, D.~Niyato, J.~Kang, D.~I. Kim, and C.~Miao, ``Optimal targeted
  advertising strategy for secure wireless edge metaverse,'' \emph{arXiv
  preprint arXiv:2111.00511}, 2021.

\bibitem{bozorgchenani2021computation}
A.~Bozorgchenani, S.~Maghsudi, D.~Tarchi, and E.~Hossain, ``Computation
  offloading in heterogeneous vehicular edge networks: On-line and off-policy
  bandit solutions,'' \emph{IEEE Transactions on Mobile Computing}, 2021.

\bibitem{yang2007fast}
Y.~Yang, P.~Yuhua, and L.~Zhaoguang, ``A fast algorithm for ycbcr to rgb
  conversion,'' \emph{IEEE Transactions on Consumer Electronics}, vol.~53,
  no.~4, pp. 1490--1493, 2007.

\bibitem{eyerman2011fine}
S.~Eyerman and L.~Eeckhout, ``Fine-grained dvfs using on-chip regulators,''
  \emph{ACM Transactions on Architecture and Code Optimization (TACO)}, vol.~8,
  no.~1, pp. 1--24, 2011.

\bibitem{wang2019energy}
H.~Wang, B.~Kim, J.~Xie, and Z.~Han, ``How is energy consumed in smartphone
  deep learning apps? executing locally vs. remotely,'' in \emph{2019 IEEE
  Global Communications Conference (GLOBECOM)}.\hskip 1em plus 0.5em minus
  0.4em\relax IEEE, 2019, pp. 1--6.

\bibitem{zeng2020volunteer}
F.~Zeng, Q.~Chen, L.~Meng, and J.~Wu, ``Volunteer assisted collaborative
  offloading and resource allocation in vehicular edge computing,'' \emph{IEEE
  Transactions on Intelligent Transportation Systems}, vol.~22, no.~6, pp.
  3247--3257, 2020.

\bibitem{wu2020fog}
Y.~Wu, J.~Wu, L.~Chen, G.~Zhou, and J.~Yan, ``Fog computing model and efficient
  algorithms for directional vehicle mobility in vehicular network,''
  \emph{IEEE Transactions on Intelligent Transportation Systems}, vol.~22,
  no.~5, pp. 2599--2614, 2020.

\bibitem{jiao2020toward}
Y.~Jiao, P.~Wang, D.~Niyato, B.~Lin, and D.~I. Kim, ``Toward an automated
  auction framework for wireless federated learning services market,''
  \emph{IEEE Transactions on Mobile Computing}, vol.~20, no.~10, pp.
  3034--3048, 2020.

\bibitem{singh2022interplay}
K.~Singh and S.~Mishra, ``Interplay of data in digital economy and merger
  control regime: A conundrum without solutions,'' \emph{Law, State and
  Telecommunications Review}, vol.~14, no.~2, pp. 17--37, 2022.

\bibitem{yang2012crowdsourcing}
D.~Yang, G.~Xue, X.~Fang, and J.~Tang, ``Crowdsourcing to smartphones:
  Incentive mechanism design for mobile phone sensing,'' in \emph{Proceedings
  of the 18th annual international conference on Mobile computing and
  networking}, 2012, pp. 173--184.

\bibitem{feng2021min}
J.~Feng, L.~Liu, Q.~Pei, and K.~Li, ``Min-max cost optimization for efficient
  hierarchical federated learning in wireless edge networks,'' \emph{IEEE
  Transactions on Parallel and Distributed Systems}, 2021.

\bibitem{liu2018edge}
Q.~Liu, S.~Huang, J.~Opadere, and T.~Han, ``An edge network orchestrator for
  mobile augmented reality,'' in \emph{IEEE INFOCOM 2018-IEEE Conference on
  Computer Communications}.\hskip 1em plus 0.5em minus 0.4em\relax IEEE, 2018,
  pp. 756--764.

\bibitem{belotti2013mixed}
P.~Belotti, C.~Kirches, S.~Leyffer, J.~Linderoth, J.~Luedtke, and A.~Mahajan,
  ``Mixed-integer nonlinear optimization,'' \emph{Acta Numerica}, vol.~22, pp.
  1--131, 2013.

\bibitem{boyd2003subgradient}
S.~Boyd, L.~Xiao, and A.~Mutapcic, ``Subgradient methods,'' \emph{lecture notes
  of EE392o, Stanford University, Autumn Quarter}, vol. 2004, pp. 2004--2005,
  2003.

\bibitem{boyd2004convex}
S.~Boyd, S.~P. Boyd, and L.~Vandenberghe, \emph{Convex optimization}.\hskip 1em
  plus 0.5em minus 0.4em\relax Cambridge university press, 2004.

\bibitem{9745059}
J.~Feng, W.~Zhang, Q.~Pei, J.~Wu, and X.~Lin, ``Heterogeneous computation and
  resource allocation for wireless powered federated edge learning systems,''
  \emph{IEEE Transactions on Communications}, pp. 1--1, 2022.

\bibitem{li2018wirelessly}
X.~Li, C.~You, S.~Andreev, Y.~Gong, and K.~Huang, ``Wirelessly powered crowd
  sensing: Joint power transfer, sensing, compression, and transmission,''
  \emph{IEEE Journal on Selected Areas in Communications}, vol.~37, no.~2, pp.
  391--406, 2018.

\bibitem{feng2019cooperative}
J.~Feng, F.~R. Yu, Q.~Pei, X.~Chu, J.~Du, and L.~Zhu, ``Cooperative computation
  offloading and resource allocation for blockchain-enabled mobile-edge
  computing: A deep reinforcement learning approach,'' \emph{IEEE Internet of
  Things Journal}, vol.~7, no.~7, pp. 6214--6228, 2019.

\bibitem{mao2017stochastic}
Y.~Mao, J.~Zhang, S.~Song, and K.~B. Letaief, ``Stochastic joint radio and
  computational resource management for multi-user mobile-edge computing
  systems,'' \emph{IEEE Transactions on Wireless Communications}, vol.~16,
  no.~9, pp. 5994--6009, 2017.

\bibitem{ahn2019novel}
J.~Ahn, J.~Lee, S.~Yoon, and J.~K. Choi, ``A novel resolution and power control
  scheme for energy-efficient mobile augmented reality applications in mobile
  edge computing,'' \emph{IEEE Wireless Communications Letters}, vol.~9, no.~6,
  pp. 750--754, 2019.

\end{thebibliography}

\end{spacing}

~\vspace{-20pt}

% \begin{center}
% \textbf{APPENDIX}
% \end{center}
\begin{center}
{Appendix A. Proof of Theorem 1}\vspace{-10pt}
\end{center}
\textit{Proof:} Since $R_{n}=B_n\log_2(1+\frac{P_{n}h_{n}}{B_n\sigma^2})$ is concave with respect to $P_n$, $\frac{1}{R_n}$ is convex \cite{boyd2004convex}. Recall $F_n(P_n) \triangleq  \frac{\gamma_n\varphi s_n^2P_n}{B_n\log_2(1+\frac{P_nh_n}{B_n\sigma^2})}$.
Wherefore, we can derive that $F_n(P_n)$ is strictly quasiconvex on $P_n$ \cite{boyd2004convex}. Furthermore, from $\mathrm{(C4)}$, $\mathrm{(C6)}$, and $\mathrm{(C7)}$, the feasible region of (\ref{TP2}) is convex.  Therefore, (\ref{TP2}) is a quasiconvex optimization problem because it minimizes a quasiconvex function over a convex set \cite{boyd2004convex}.

\begin{center}
{Appendix B. Proof of Theorem 2}\vspace{-10pt}
\end{center}

Based on the definition of $D(\nu, \boldsymbol \mu)$ in Equation~(\ref{L}), with $f_n^{s*}$ denoting the optimal solution corresponding to $\nu$ and $\boldsymbol \mu$, we have
\begin{eqnarray} 
D(\nu', \boldsymbol \mu')=\min\limits_{\boldsymbol f^s\text{ satisfying }(\mathrm{C7})} L(\boldsymbol f^s, \nu', \boldsymbol \mu') \leq L(\boldsymbol f^{s*}, \nu', \boldsymbol \mu'),
\end{eqnarray}
which means
\begin{eqnarray}\label{D1}
&& \hspace{-40pt} -D(\nu', \boldsymbol \mu') \hspace{-2pt} \geq \hspace{-2pt}  -\sum\limits_{n\in\mathcal{N}}\beta_n\kappa_{ser}(f_n^{s*})^2\varphi s_n^2c_n\hspace{-2pt} -\hspace{-2pt} \nu'\left(\sum\limits_{n\in\mathcal{N}}f_n^{s*}\hspace{-2pt}-\hspace{-2pt}F\right)\hspace{-2pt} -\hspace{-2pt} \sum\limits_{n\in\mathcal{N}}\mu'_n\left(\frac{\varphi s_n^2c_n}{T-T_n^{cv}-\frac{\varphi s_n^2}{R_n}}\hspace{-2pt}-\hspace{-2pt}f_{n}^{s*}\right)\hspace{-3pt}.
\end{eqnarray}
% where the inequality holds since $f_n^{s*}$ is a optimal solution corresponding to $\nu$ and $\boldsymbol \mu$.

We rearrange (\ref{D1}) as
\begin{eqnarray}
&&\!\!\!\!\!-D(\nu', \boldsymbol \mu')\geq -D(\nu, \boldsymbol \mu)+(\nu'-\nu)\left(F-\sum\limits_{n\in\mathcal{N}}f_n^{s*}\right) +\sum\limits_{n\in\mathcal{N}}(\mu'_n-\mu_n)\left(f_{n}^{s*}-\frac{\varphi s_n^2c_n}{T-T_n^{cv}-\frac{\varphi s_n^2}{R_n}}\right). \nonumber
\end{eqnarray}

It is worth nothing that if $y(x_1)\geq y(x_2)+m^T(x_1-x_2)$ holds for $x_1$ and $x_2$ in the domain, then $m$ is defined as a subgradient of a convex function $y(\cdot)$. Therefore, we have proved Theorem 2.

\end{document}